\documentclass[journal]{IEEEtran}
\usepackage{amsmath,amsfonts}
\usepackage[linesnumbered, ruled, lined,boxed]{algorithm2e}
\usepackage{array}
\usepackage[caption=false,font=normalsize]{subfig}
\usepackage{textcomp}
\usepackage{stfloats}
\usepackage{url}
\usepackage{verbatim}
\usepackage{graphicx}
\usepackage{cite}
\usepackage{flushend}
\usepackage{amsthm}
\usepackage{amssymb}
\usepackage{physics}
\usepackage{tikz}
\usepackage{lipsum}
\usepackage{makecell}

\usepackage{xcolor}
\newtheorem{thm}{Theorem}
\newtheorem{lem}{Lemma}
\newtheorem{defi}{Definition}
\newtheorem{remark}{Remark}

\SetKwInput{KwIn}{Input} 
\SetKwInput{KwOut}{Output} 

\begin{document}

\allowdisplaybreaks

\title{
Event-Triggered Regulation of Mixed-Autonomy Traffic Under Varying Traffic Conditions}

\author{ Yihuai Zhang~\IEEEmembership{Graduate Student Member,~IEEE}, Huan Yu*~\IEEEmembership{Senior Member, IEEE}

\thanks{Yihuai Zhang and Huan Yu are with the Hong Kong University of Science and Technology (Guangzhou), Thrust of Intelligent Transportation, Guangzhou, China. This work was supported by National Natural Science Foundation of China under Project 62203131.}

\thanks{*Corresponding author (e-mail:huanyu@hkust-gz.edu.cn)}}



\maketitle

\begin{abstract}
Modeling and congestion mitigation of mixed-autonomy traffic systems consisting of human-driven vehicles (HVs) and autonomous vehicles (AVs) have become increasingly critical with the rapid development of autonomous driving technology. This paper develops an event-triggered control (ETC) framework for mitigating congestion in such systems, which are modeled using an extended Aw-Rascle-Zhang (ARZ) formulation consisting of coupled 4$\times$4 hyperbolic partial differential equations (PDEs). Ramp metering is employed as the boundary actuation mechanism. To reduce computational and communication burdens while avoiding excessive ramp signal changes, we design the ETC strategy based on the backstepping method, together with an observer-based ETC formulation for practical implementation under limited sensing. Rigorous Lyapunov analysis ensures exponential convergence and avoidance of Zeno behavior. Extensive simulations validate the proposed approach under diverse traffic scenarios, including varying AV penetration rates, different spacing policies, multiple demand levels, and non-recurrent congestion patterns. Results show that ETC not only stabilizes mixed traffic flows but also significantly reduces control updates, improving driver comfort, and roadway safety. Higher AV penetration rates lead to longer release time and fewer triggering events, indicating the positive impact of AVs in mitigating traffic congestion while reducing computational resource usage. Compared to continuous backstepping controllers, the proposed ETC achieves near-equivalent stabilization performance with far fewer controller updates, resulting in longer signal release time that reduces driver distraction, which demonstrates great potential for ETC applications in traffic management.
\end{abstract}


\section{Introduction}
\IEEEPARstart{F}{reeway} traffic congestion causes increased total travel time, risk of accidents,  and fuel consumption~\cite{siri2021freeway}. Various control strategies have been proposed to mitigate freeway traffic congestion, e.g., feedback control~\cite{papageorgiou1991alinea,karafyllis2019feedback,zhang2017stochastic}, optimal control~\cite{gomes2006optimal,goatin2016speed,delle2017traffic}, backstepping control~\cite{yu2019traffic}, and learning-based control~\cite{belletti2017expert}. In this paper, we mainly focus on the macroscopic freeway traffic and the design of infrastructure-based control strategies for ramp metering or varying speed limits. As autonomous driving technology advances rapidly, the increasing prevalence of autonomous vehicles~(AVs) has created mixed-autonomy traffic systems where human-driven vehicles~(HVs) and AVs share roadways. The dynamic interactions between these two vehicle types often cause traffic flow instabilities, while the development of controller design for such heterogeneous traffic environments remains a significant challenge in transportation research.
\subsection{Mixed-autonomy traffic modeling and control}
Freeway traffic dynamics are described by partial differential equations~(PDEs) using aggregated traffic states, e.g., Aw-Rascle-Zhang~(ARZ) model~\cite{aw2000resurrection,zhang2002non}.
The mixed-autonomy traffic can be modeled as a type of multi-class traffic where AVs and HVs adopt different spacing policies. Both First-order and second-order multi-class traffic flow models capture the behaviors of different vehicle types, where slower vehicles act as moving bottlenecks for faster ones, influencing their speeds and interactions even in congested regimes~\cite{tang2009new}. In addition, Area Occupancy~(AO) which distinguishes the heterogeneous traffic participants for the multi-class traffic model was proposed in~\cite{mohan2017heterogeneous}.   {The area occupancy is defined as the vehicle length times vehicle width to measure how much area each type of vehicles occupied on the road.} In this paper, we adopt the same concept from the multi-class traffic model~\cite{zhang2023mean,burkhardt2021stop} and further consider different driving behaviors of HVs and AVs to introduce the mixed-autonomy traffic PDE model. Mixed-autonomy characteristics can also be captured using fundamental diagrams, such as Greenshields' relation between density and speed, with studies suggesting AVs may increase capacity via reduced spacing\cite{zhou2020modeling} but others arguing AVs could require larger spacing due to safety and technological limitations,   {vehicle manufactures have not yet adopted to deploy short driving spacing in commercial vehicles due to the safety concerns in the industry practice. They may yield a larger driving spacing than an average theoretical calculations~\cite{li2022trade}}. 

Various control methods have been applied for the suppression of traffic congestion on freeways by changing control signals of traffic lights on ramps and variable speed limits. Feedback controller can be applied for both continuous models~\cite{karafyllis2019feedback,zhang2019pi} and discretized models~\cite{papageorgiou1991alinea}. In addition, optimal control strategies address distinct objectives: minimizing delays~\cite{gomes2006optimal} or reducing outflow errors~\cite{delle2017traffic}. Among these control methods, backstepping control is a widely-adopted design method for PDE control. It does not require numerical discretization in design and achieves the finite-time convergence and stability guarantees~\cite{krstic2008boundary}. The PDE backstepping was firstly adopted ~\cite{yu2019traffic} and systematically applied to traffic flow models by~\cite{yu2022traffic}. 
  {
Our previous work analyzed the stability of the mixed-autonomy traffic under stochastic effect using {backstepping}~\cite{zhang2023mean} and developed backstepping controller to achieve mean-square exponential stability. However, the computational burden caused by calculating {backstepping} control signals that are continuous in time can be high. It also raises practical issues for implementation, since controlled vehicles are forced to continuously change their speed by obeying the traffic lights or speed limits. 
}
  {Therefore, developing efficient control methods for traffic system is essential, not only improving the efficiency of the traffic system but also reducing the cost for the practical implementation.}

\subsection{Event-triggered control}
  {Event-triggered control (ETC) is a control design strategy aimed at improving system efficiency by updating the control input only when certain triggering conditions are satisfied. It requires the definition of a triggering mechanism to determine the instants at which the controller must be updated. The triggering condition can be either static or dynamic~\cite{girard2014dynamic,tabuada2007event,wang2024event}. The first ETC design for 2 $\times$ 2 hyperbolic PDEs was proposed in~\cite{espitia_event-based_2018}, and later extended to observer-based output-feedback control in~\cite{espitia2020observer}. Typically, a continuous-time control law is first designed to stabilize the system; the event-triggered control input—guaranteeing the stability properties—is then derived based on the triggering condition. While the works in~\cite{espitia_event-based_2018,espitia2020observer} established foundational ETC and observer-based ETC frameworks for 2 $\times$ 2 hyperbolic PDEs, including rigorous proofs of exponential stability, they focus primarily on theoretical developments and do not address traffic-specific applications or the robustness of ETC under varying traffic conditions. In the context of traffic systems, Espitia et al.~\cite{espitia2022traffic} developed an output-feedback ETC for cascaded freeway segments using variable speed limits. The system was modeled as a 4 $\times$ 4 hyperbolic PDE system, and exponential convergence of the closed-loop system under the proposed ETC was proven. To further reduce the number of control updates, a performance-barrier-based ETC was recently proposed in~\cite{zhang2024performance}, achieving longer dwell times and enhancing driver safety.}

  {However, existing continuous-flow PDE models have not yet considered mixed-autonomy traffic scenarios, focusing instead on homogeneous (pure HVs) traffic. ETC has also been applied to discretized traffic models, often in combination with other control strategies such as model predictive control (MPC) or local feedback controllers~\cite{ferrara2015event,pasquale2020hierarchical}. While these approaches can effectively minimize performance metrics such as queue length and total travel time, they generally lack rigorous stability guarantees for the traffic system.}

  {Furthermore, in all ETC designs, the Zeno phenomenon must be explicitly ruled out. The Zeno phenomenon refers to the occurrence of infinitely many triggering events within a finite time interval, which can lead to excessive computational load and potentially destabilize the control system. In practical traffic applications, Zeno behavior would severely hinder the implementability of the control input due to unrealistic communication and actuation demands. In this paper, we develop an ETC framework for mixed-autonomy traffic systems that provides theoretical stability guarantees while also demonstrating practical applicability in realistic traffic scenarios.}

\subsection{Contribution}
  {
Previous efforts to mitigate traffic congestion have primarily focused on homogeneous traffic consisting solely of HVs. Although observer-based ETC designs have been proposed for cascaded traffic systems~\cite{espitia2022traffic}, results on ETC for mixed-autonomy traffic modeled by PDEs remain limited.
}

  {
Building upon our preliminary theoretical work~\cite{zhang2024event,zhang2023mean} and the event-triggered control design for 2 $\times$ 2 hyperbolic PDEs~\cite{espitia_event-based_2018}, we first establish the congestion analysis for the mixed-autonomy traffic system using the extended ARZ model, and identify the free and congested regimes under different equilibrium density settings. From an application perspective, we further develop and analyze event-triggered control strategies under various traffic conditions—including different demand levels and AV penetration rates—and evaluate their impact on overall traffic performance. Moreover, we demonstrate that the proposed ETC framework significantly reduces the computational burden on traffic management systems, while simultaneously improving traffic efficiency and safety. This enables scalable deployment in large-scale transportation networks.
}

The structure of the paper is as follows: Section~\ref{sec2} develops the mixed-autonomy traffic PDE model and analyzes the free/congestion conditions for the model. Section~\ref{sec3} gives the continuous stabilizing controller for the traffic system using the backstepping method. Section~\ref{sec4} details the design of the event-triggered controller and gives the theoretical analysis for the system to avoid the Zeno phenomenon and finite time convergence. Section~\ref{sec5} conducts extensive experiments to test the proposed ETC under different traffic conditions. Finally, Section~\ref{sec6} concludes the paper.

\section{Mixed-Autonomy Traffic PDE Model} \label{sec2}
In this section, we introduce the mixed-autonomy traffic flow model that explicitly accounts for the distinct driving behaviors of HVs and AVs~\cite{zhang2024event},
\begin{align}
    \partial_t \rho_{\rm h}+\partial_x\left(\rho_{\rm h} v_{\rm h}\right) &=0, \\
    \partial_t\left(v_{\rm h}-V_{e,\rm h}\right)+v_{\rm h} \partial_x\left(v_{\rm h} -V_{e,\rm h}(AO) \right)&=\frac{V_{e, \rm h}(AO)-v_{\rm h}}{\tau_{\rm h}}, \\
    \partial_t \rho_{\rm a}+\partial_x\left(\rho_{\rm a} v_{\rm a}\right) &=0,\\
    \partial_t\left(v_{\rm a}-V_{e, \rm a}(AO)\right)+v_{\rm a} \partial_x \left(v_{\rm a}-V_{e, \rm a}\right) &=\frac{V_{e, \rm a}(AO)-v_{\rm a}}{\tau_{\rm a}},
\end{align}
where ${\rho}_{\rm h}(x,t)$, ${\rho}_{\rm a}(x,t)$, ${v}_{\rm h}(x,t)$, and ${v}_{\rm a}(x,t)$ denote traffic densities and speeds of HVs and AVs. With the following boundary conditions:
\begin{align}
    \rho_{\rm h}(0, t) & =\rho_{\rm h}^\star, \\
    \rho_{\rm a}(0, t) & =\rho_{\rm a}^\star, \\
    \rho_{\rm h}(0, t) v_{\rm h}(0, t)+\rho_{\rm a}(0, t) v_{\rm a}(0, t) & =\rho_{\rm h}^\star v_{\rm h}^\star+\rho_{\rm a}^\star v_{\rm a}^\star, \\
    \rho_{\rm h}(L, t) v_{\rm h}(L, t)+\rho_{\rm a}(L, t) v_{\rm a}(L, t) & =q_{\rm h}^\star+q_{\rm a}^\star + U(t),
\end{align}
where the traffic states are defined in the spatial and time domain $(x,t) \in [0,L] \times \mathbb{R}^+$,  $\rho_{\rm h}^\star$, $\rho_{\rm a}^\star$ denote equilibrium densities, while $v_{\rm h}^\star$, $v_{\rm a}^\star$ are the equilibrium speeds. The equilibrium flow is calculated by $q^\star_{\rm h} = \rho_{\rm h}^\star \times v_{\rm h}^\star$, $q^\star_{\rm a} = \rho_{\rm a}^\star \times v_{\rm a}^\star$. The $\tau_{\rm h}$ and $\tau_{\rm a }$ are the reaction times for HVs and AVs, describing how long the drivers can adapt to the desired speed. The $V_{e,\rm h}(AO)$ and $V_{e, \rm a}(AO)$ denote the fundamental diagram for HVs and AVs, describing the density-speed relation of the traffic system. We adopt another measurement, $AO$, to describe the mixed density for the mixed-autonomy traffic system.

\begin{figure}
    \centering
    \includegraphics[width=0.9\linewidth]{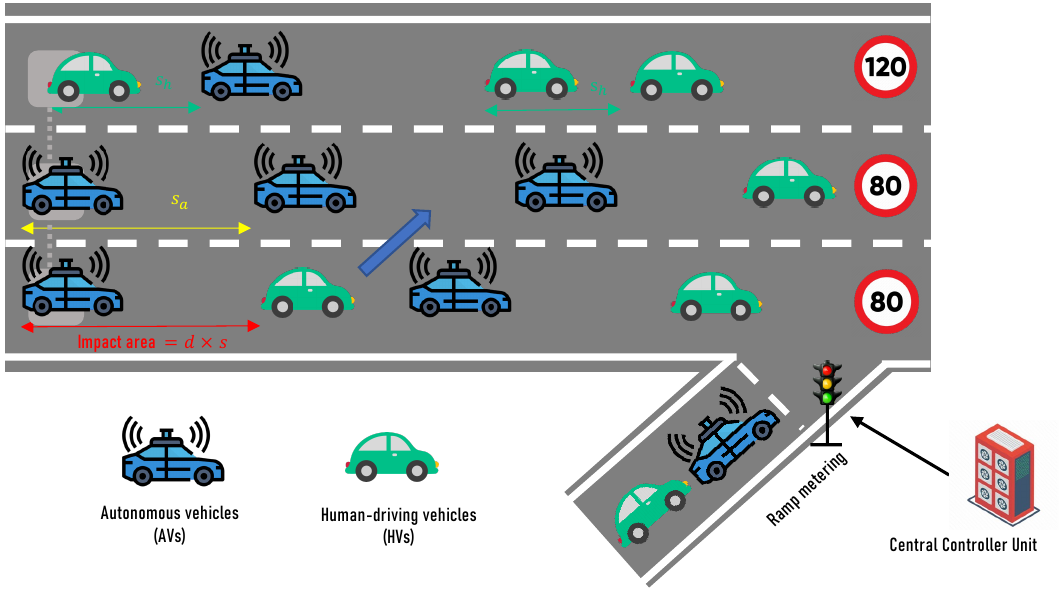}
    \caption{The mixed-autonomy traffic system with different spacing policy. $s_{\rm h}$ denotes the spacing of HVs, $s_{\rm a}$ represents the spacing of AVs. The impact area is calculated by Eq.~\eqref{impactHV} and~\eqref{impactAV}.}
    \label{mixed}
\end{figure}

The first two equations describe the density and speed evolution of HVs on the road section, while the other two equations denote the evolution of AVs. For the boundary conditions for HVs and AVs, we assume the inlet density at its equilibrium density for both HVs and AVs.   {In addition, the upstream flow and the downstream flow also stay at the equilibrium flow $q^\star_{\rm h} + q^\star_{\rm a}$.} The boundary conditions are assumed for an averaged short time period of the road, therefore, the penetration rate of HVs and AVs remains the same and the traffic conditions do not significantly change during the short time period.   {The control input $U(t)$ is the flow perturbation around the equilibrium flow and is implemented by the ramp metering, which needs to be designed to stabilize the mixed-autonomy traffic system.} The schematic diagram for the mixed-autonomy traffic system is shown in Fig.~\ref{mixed}.
\subsection{Different spacing policies and fundamental diagram}
\begin{figure}
    \centering
    \subfloat[$V_{e,\rm h}(AO)$]{\includegraphics[width=0.45\linewidth]{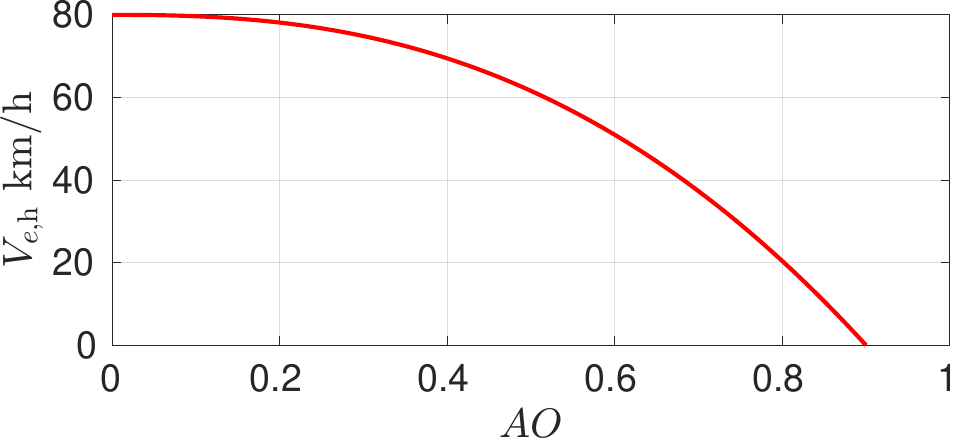}}
    \subfloat[$V_{e,\rm a}(AO)$]{\includegraphics[width=0.45\linewidth]{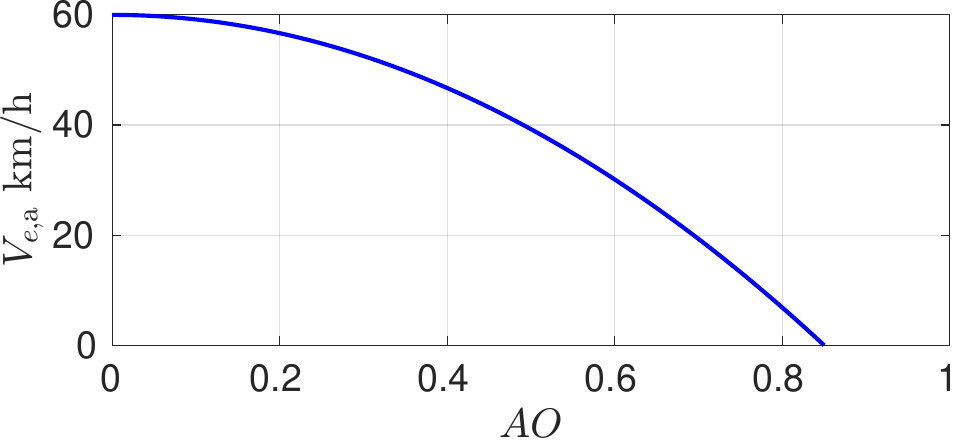}}
    \caption{The fundamental diagram of HVs and AVs using area occupancy(AO)}
    \label{fundiaAO}
\end{figure}

\begin{figure}
    \centering
    \subfloat[$V_{e,\rm h}(\rho_{\rm h},\rho_{\rm a})$]{\includegraphics[width=0.45\linewidth]{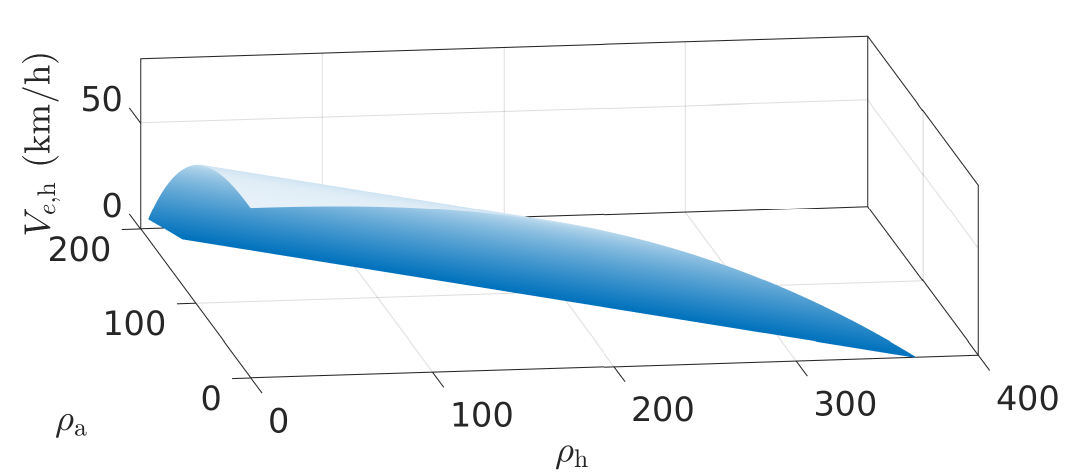}}
    \subfloat[$V_{e,\rm a}(\rho_{\rm h},\rho_{\rm a})$]{\includegraphics[width=0.45\linewidth]{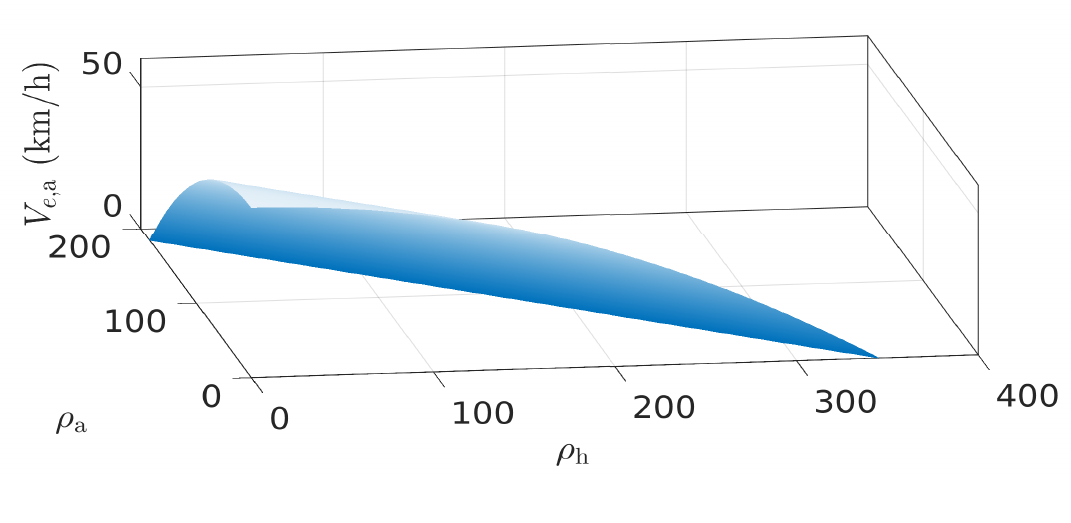}}
    \caption{The fundamental diagram of HVs and AVs using original densities}
    \label{fundiarho}
\end{figure}

  {We use $AO$ to describe the interaction between different types of vehicles on the road~\cite{zhang2024event,burkhardt2021stop,mohan2017heterogeneous}. We adopt this concept containing microscopic traffic parameters(vehicle width and spacing) to distinguish driver behaviors of different types of vehicles in the mixed-autonomy traffic system. Then the area occupancy is incorporated in the fundamental diagram $V_{e,h}(AO)$, $V_{e,a}(AO)$ of the macroscopic traffic model to describe the interaction of AVs and HVs. The definition of $AO$ is given by:}
\begin{align}
    AO(\rho_{\rm h},\rho_{\rm a}) = \frac{a_{\rm h} \rho_{\rm h} + a_{\rm a} \rho_{\rm a}}{W}, \label{AOrelation}
\end{align}
where $W$ denotes the road width. $a_{\rm h}$ and $a_{\rm a}$ are impact areas, defined as: 
\begin{align}
    a_{\rm h} &= d \times s_{\rm h} \label{impactHV},\\
    a_{\rm a} &= d \times s_{\rm a} \label{impactAV}.
\end{align}
$d$ denotes the vehicle width. In this paper, the vehicle width is assumed to be the same. $s_{\rm h}$, $s_{\rm a}$  are the spacing policy of HVs and AVs. Compared with HVs, AVs tend to have a larger spacing due to the safety issue caused by the current state of autonomous driving level~\cite{li2022trade}. When AVs maintain larger following gaps, the resultant spacing creates a large impact area that enables HVs to interleave with AVs through ``creeping behavior'' during traffic congestion. To characterize the equilibrium relationship between traffic speed and density, the fundamental diagram based on roadway occupancy metrics is expressed as:
\begin{align}
    v_{\rm h} = V_{e,\rm h}(\rho_{\rm h},\rho_{\rm a}) = V_{\rm h} \left( 1 - \left(\frac{AO}{\overline{AO}_{\rm h}}\right)^{\gamma_{\rm h}}\right),\\
    v_{\rm a} = V_{e,\rm a}(\rho_{\rm h},\rho_{\rm a}) = V_{\rm a} \left( 1 - \left(\frac{AO}{\overline{AO}_{\rm a}}\right)
    ^{\gamma_{\rm a}}\right),
\end{align}
where $V_{\rm h}$, $V_{\rm a}$ are the free-flow speeds of HVs and AVs, $\overline{AO}_{\rm h}$, $\overline{AO}_{\rm a}$ are the maximum area occupancy for each type of vehicles, $\gamma_{\rm h}$, $\gamma_{\rm a}$ denote the traffic pressure exponent. It is often noted that commercial AVs typically maintain a greater following distance compared to HVs. This is primarily due to safety considerations, as AVs require an adequate buffer to respond effectively in potentially hazardous situations, i.e., a sudden stop by the vehicle ahead~\cite{li2022trade}.
Using the parameter setting in the experiment part, we get the fundamental diagram shown in Fig.~\ref{fundiaAO}.
As in~\eqref{AOrelation}, the $AO$ consists of the density of AVs and HVs. When we apply the original density to the fundamental diagram, we get the results shown in Fig.~\ref{fundiarho}. The fundamental diagram of HVs and AVs describes the relation between the speed and the density of HVs and AVs. This also coincides with the fact that $AO$ could reflect the interaction of the two types of vehicles. 
  {
\begin{remark}
    In our setting, we mainly focus on the general one-lane traffic, also called the lane-free traffic. There are no dished lines to divide the road into parallel lanes and faster vehicles can overtake slower ones. vehicles are not bound to fixed traffic lanes as in conventional traffic, they may derive anywhere on the 2-D surface of the road, respecting the road boundaries. Therefore,  the lane-free traffic would have larger lateral area occupancy, as well as the maximum area occupancy $\overline{AO}_{\rm h}$, $\overline{AO}_{\rm a}$. The fundamental diagram of AVs and HVs would change and the traffic capacity would increase in the lane-free traffic scenario~\cite{papageorgiou_lane-free_2021}.
\end{remark}
}

\subsection{Free/congested traffic analysis}
The model we developed are $4 \times 4$ nonlinear PDEs. We will analyze the free and congested traffic and then design boundary controller by first taking linearization of the model at its equilibrium point $\rho_{\rm h}^\star$, $\rho_{\rm a}^\star$, $v_{\rm h}^\star$, $v_{\rm a}^\star$. We define small deviations between the traffic states and equilibrium states $\Tilde{\rho}_{\rm h}(x,t) = \rho_{\rm h}(x,t) - \rho_{\rm h}^\star$, $\Tilde{v}_{\rm h}(x,t) = v_{\rm h}(x,t) - v_{\rm h}^\star$, $\Tilde{\rho}_{\rm a}(x,t) = \rho_{\rm a}(x,t) - \rho_{\rm a}^\star$, $\Tilde{v}_{\rm a}(x,t) = v_{\rm a}(x,t) - v_{\rm a}^\star$.   {The system can be rewritten using an augmented expression $\mathbf{z}(x,t) = \begin{bmatrix}
    \Tilde{\rho}_{\rm h}(x,t)  &  \Tilde{v}_{\rm h}(x,t) & \Tilde{\rho}_{\rm a}(x,t) & \Tilde{v}_{\rm a}(x,t)
\end{bmatrix}^\mathsf{T}$.} 
Defining the matrix $\mathbf{V} =\{{\nu}_{ij}\}_{1 \leq i, j \leq 4}$ such that the coefficient matrix can be diagonalized as $\mathbf{V}^{-1} \mathbf{J}_{\lambda} \mathbf{V}$ = $\text{Diag}\{\lambda_1,\lambda_2,\lambda_3,\lambda_4\}$, with  positive eigenvalues in ascending order. We also define the source term matrix as $\Hat{\mathbf{J}} = \mathbf{V}^{-1}{\mathbf{J}} \mathbf{V} = \{ \Hat{J}_{ij} \}_{1 \leq i,j \leq 4}$. 
The change of coordinates is
\begin{align}
\begin{bmatrix}
    w_1 & w_2 & w_3 & w_4
\end{bmatrix}^\mathsf{T}=\mathbf{T} \mathbf{z}. \label{transfomation}
\end{align}
where $\mathbf{T}$ is the same transformation matrix in our previous results~\cite{zhang2023mean}.
Using the transformation $\mathbf{T}$, thus we get the following boundary control model:
\begin{align}
        \mathbf{w}^+_t(x,t) +\Lambda^{+} \mathbf{w}^+_x(x,t) =& \Sigma^{++}(x)\mathbf{w}^+ (x,t)\nonumber\\
        &+\Sigma^{+-}(x) \mathbf{w}^-(x, t), \label{clpsys1}\\
        \mathbf{w}^-_t(x, t)-\Lambda^{-} \mathbf{w}^-_x(x, t) =& \Sigma^{-+}(x) \mathbf{w}^+(x,t),\label{clpsys2}\\
    \mathbf{w}^+(0,t)  =& {Q} \mathbf{w}^-(0, t), \label{clpsys3} \\
    \mathbf{w}^-(L, t) =& {R}\mathbf{w}^+(L, t)+\bar{U}(t), \label{clpsys4}
\end{align}
where $\mathbf{w}^{+} = [w_1, w_2, w_3]^\mathsf{T}$, $\mathbf{w}^{-} = w_4$. The coefficient matrices are given as
$\Lambda^{+} =\text{Diag}\{\lambda_{1},\lambda_{2},\lambda_{3}\}$, $\Lambda^- = -\lambda_4$. $\Sigma^{++}(x)$, $\Sigma^{+-}(x)$, $\Sigma^{-+}(x)$, $Q$, and $R$ are corresponding coefficients which are detailed in~\cite{burkhardt2021stop,zhang2023mean}. Also, $\bar{U}(t)=\mathrm{e}^{-\frac{\Hat{J}_{44} }{\lambda_{4}} L} \frac{1}{\kappa_{4}} U(t)$, $\kappa_{j}=v_{\rm h}^\star \nu_{1 j} +\rho_{\rm h}^\star \nu_{2 j} +v_{\rm a}^\star \nu_{3 j} +\rho_{\rm a}^\star \nu_{4 j} , j=1,2,3,4$.
It was shown that the eigenvalues satisfying the following condition \cite{zhang2006hyperbolicity}
\begin{align}
     \lambda_{4} \leq \min\{ \lambda_{1},\lambda_{3} \} \leq \lambda_{2}\leq\max\{\lambda_{1},\lambda_{3}\}.
\end{align}
Traffic flow regimes can be differentiated into free-flow and congested states through analysis of traffic wave propagation direction.
\begin{itemize}
    \item Free-flow region: $\lambda_{1}>0$, $\lambda_{2}>0$, $\lambda_{3}>0$, {$\lambda_{4}>0$}. Traffic oscillations transport downstream at corresponding speed $\lambda_{1}$, $\lambda_{2}$, $\lambda_{3}$, $\lambda_{4}$. Vehicles run at their maximum speeds.
    \item Congested region: $\lambda_{1}>0$, $\lambda_{2}>0$, $\lambda_{3}>0$, {$\lambda_{4}<0$}. Traffic information propagates from downstream to upstream triggers efficiency degradation in transportation systems.
\end{itemize}
In this paper, we focus on the congested region for the mixed-autonomy traffic system, which can be determined by the value of the characteristic speed $\lambda_4$.   {When $\lambda_4<0 \rightarrow \textit{negative wave speed}$}, the traffic system lies in the congested region. 

\begin{figure}
    \centering
    \includegraphics[width=0.9\linewidth]{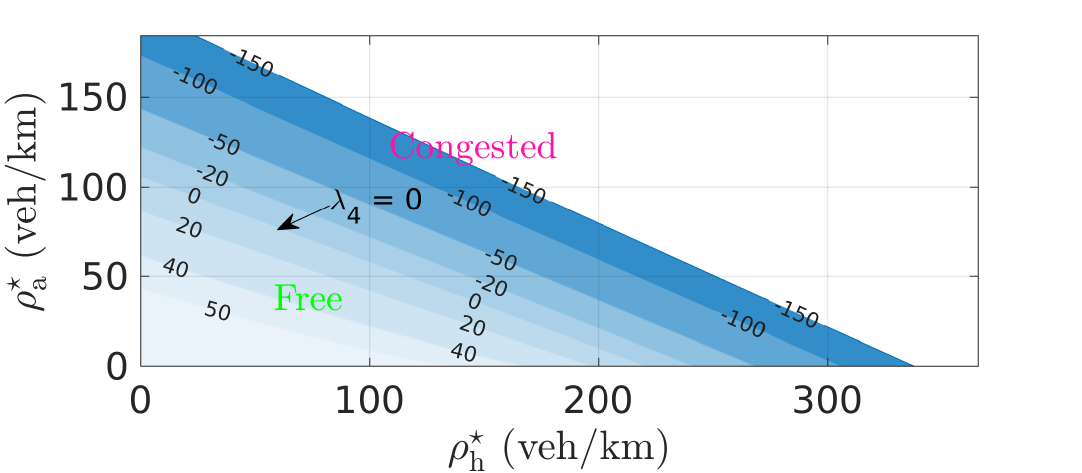}
    \caption{The free and congestion region determined by $\lambda_4$. $\lambda_4<0$ denotes the congested region while the other denotes the free region of the mixed-autonomy traffic system.}
    \label{congestedregion}
\end{figure}

The value of $\lambda_4$ is related to the equilibrium density of AVs and HVs. We give the free/congested region related to the different equilibrium densities, as shown in Fig.~\ref{congestedregion}. In this paper, we mainly focus on the congested region of the mixed-traffic system. It also guides us in the selection of the system parameters in the experiment part.

\section{Boundary Controller Design} \label{sec3}
In this section, we propose the boundary controller using the backstepping method. The backstepping controller could stabilize the traffic system in finite time such that all traffic states converge to their equilibrium points. The event-trigger strategies will be developed based on the backstepping design.

\subsection{Backstepping transformation}
Define the backstepping transformation that transforms the original system \eqref{clpsys1}-\eqref{clpsys4} into the target system:
\begin{align}
    \mathcal{K}\mathbf{w}= \begin{pmatrix} \mathbf{w}^{+} \\
    \mathbf{w}^- -\int_0^x \mathbf{K}(x, \xi)\mathbf{w}^{+}(\xi,t) +M(x,\xi) \mathbf{w}^-(\xi, t) d \xi \end{pmatrix},\label{back4}
\end{align}
where $\mathbf{w} = [\mathbf{w}^{+}, \mathbf{w}^{-}]$ and the backstepping control kernels $\mathbf{K}(x,\xi)\in \mathbb{R}^{3}$, $M(x,\xi) \in \mathbb{R}^{1}$ are:
\begin{align}
    \mathbf{K}(x, \xi)=\left[\begin{array}{lll}
    k_{1}(x, \xi) & k_{2}(x, \xi) & k_{3}(x, \xi)
    \end{array}\right]
\end{align}
The backstepping kernels are defined on the triangular domain $\mathcal{T}=\{0 \leq \xi \leq x \leq L\}$. 
We get the target system through the backstepping transformation:
\begin{align}
\alpha_t(x,t)+\Lambda^{+} \alpha_x(x,t) =&\Sigma^{++}(x)\alpha(x,t) +\Sigma^{+-}(x) \beta(x,t) \nonumber \\
+\int_0^x\mathbf{C}^+(x,\xi)\alpha(\xi,t) d\xi &+ \int_0^x\mathbf{C}^-(x,\xi)\beta(\xi,t)d\xi,
\label{tarsys1}\\
\beta_t\left(x, t\right)-\Lambda^{-} \beta_x(x, t)&=0, \label{tarsys2}\\
    \alpha(0,t)  &= {Q} \beta(0, t), \label{tarsys3} \\
    \beta(L, t) &= 0, \label{tarsys4}
\end{align}
where $\alpha = [\alpha_1, \alpha_2, \alpha_3]^\mathsf{T}$ The coefficients $ \mathbf{C}^+(x,\xi)\in \mathbb{R}^{3\times 3}$ and $\mathbf{C}^-(x,\xi)\in \mathbb{R}^{3\times1}$ are also defined on $\mathcal{T}$. 
The kernel equations and their well-posedness, as well as the well-posedness of the target system, are provided in~\cite{zhang2024event,hu2015control}.
\subsection{Inverse Transformation}
The transformation~\eqref{back4} establishes a one-to-one correspondence between the target and original systems, preserving their structural equivalence. Conversely, applying an inverse transformation mathematically maps the target system back to the original system, defined as:
\begin{align}\label{invback}
    \mathcal{L}\vartheta  = \begin{pmatrix}
        \alpha\\
        \beta - \int_0^x (\mathbf{L}(x,\xi) \alpha(\xi,t) + N(x,\xi)\beta(\xi,t))d\xi
    \end{pmatrix},
\end{align}
where $\alpha = [\alpha_1, \alpha_2, \alpha_3]^\mathsf{T}$ are the states of the target system, $\vartheta = [\alpha, \beta]^\mathsf{T}$ and $\mathbf{L}(x,\xi) \in \mathbb{R}^3$, $N(x,\xi)\in \mathbb{R}^1$ are defined as 
\begin{align}
    \mathbf{L}(x,\xi) = \begin{bmatrix}
        \ell_1(x,\xi) & \ell_2(x,\xi) & \ell_3(x,\xi)
    \end{bmatrix}.
\end{align}
They are also defined on $\mathcal{T}$. The states $\mathbf{w}$ and $\vartheta$ have equivalent $L_2$ norms, i.e., there exist two constants $p_1>0$ and $p_2>0$ such that
 \begin{align}
    p_1\norm{\mathbf{w}}_{L^2}^2 &\leq \norm{\vartheta}_{L^2}^2\leq p_2\norm{\mathbf{w}}_{L^2}^2,
\end{align}
where $\vartheta = (\alpha_1,\alpha_2,\alpha_3,\beta)$.

The continuous-time control input $\bar{U}(t)$ is constructed through real-time state feedback from the target system's states:
\begin{align}
    \bar{U}(t)=&\int_0^L (\mathbf{L}(L, \xi)\alpha(\xi,t) +N(L, \xi) \beta (\xi, t)) {d} \xi -R\mathbf{w}^+(L,t).
    \label{control_law}
\end{align}
Based on the backstepping transformation, we have the following theorem:
\begin{thm}[\cite{burkhardt2021stop,burkhardt_suppression_2020}]
    Consider the plant~\eqref{clpsys1}-\eqref{clpsys4} with the backstepping control law~\eqref{control_law}. For given initial conditions $(\mathbf{w}^+_0,\mathbf{w}^-_0)\in L^2((0,L),\mathbb{R}^4)$, the equilibrium $\mathbf{w}\equiv 0$ is finite-time stable in the $L^2$ sense and the equilibrium is reached in finite time $t_f$ = $\frac{L}{\min{\{\lambda_1,\lambda_3\}}}$ + $ \frac{L}{-\lambda_4}$.
\end{thm}
  {The finite-time convergence of the $\mathbf{w}$ system to zero implies that small perturbations in the traffic states also vanish in finite time. A detailed proof is provided in~\cite{burkhardt_suppression_2020,burkhardt2021stop}.}

\section{Event-triggered Controller Design}\label{sec4}
This section proposes event-triggered control strategies by identifying the triggering mechanism, i.e., determining the time intervals for controller updates while guaranteeing exponential stability of the closed-loop system.

We start from event-based sampling of the boundary control input $\Bar{U}(t)$, where controller updates occur at discrete temporal intervals satisfying a predefined triggering condition. The boundary condition in~\eqref{clpsys4} is then reformulated to accommodate event-triggered updates:
\begin{align}\label{ETCboundary}
    \mathbf{w}^-(L, t) = {R}\mathbf{w}^+(L, t)+\bar{U}_d(t),
\end{align}
where $\bar{U}_d(t) = \bar{U}(t) + d(t)$, $\forall t \in [t_k,t_{k+1}),k>0$.
Here, $d(t)$ denotes the real-time discrepancy between the theoretical control input and its event-triggered implementation. The sampled-data control law is then formulated as:
\begin{align}
     \bar{U}_d(t) = &\int_0^L (\mathbf{L}(L, \xi)\alpha(\xi,t_k) +N(L, \xi) \beta (\xi, t_k)) {d} \xi \nonumber\\
     &-R\mathbf{w}^+(L,t_k), \label{control law sam}
\end{align}
the actuation discrepancy $d(t)$ is then obtained as
\begin{align}
    d(t) = &-R(\mathbf{w}^+(L,t_k) - \mathbf{w}^+(L,t)) \nonumber\\ 
    &+ \int_0^L \bigg(\mathbf{L}(L, \xi)(\alpha(\xi,t_k)-\alpha(\xi,t))\nonumber\\
    &+N(L, \xi) (\beta (\xi, t_k)- \beta (\xi, t)\bigg) {d} \xi. \label{dt}
\end{align}
  {Since $\alpha(x,t)$, $\beta(x,t)$, $\mathbf{w}^+(L,t)$ are continuous with respect to time due to the property of the mixed-autonomy traffic system, and the integration of backstepping kernels and states preserves the continuity, then $d(t)$ is $\mathcal{C}^0([t_k,t_{k+1}],\mathbb{R})$ between two triggering events.} Applying the sampled control law $\Bar{U}_d(t)$ within the system \eqref{clpsys1}-\eqref{clpsys4}, induces the perturbed target system
\begin{align}
\alpha_t(x,t)+\Lambda^{+} \alpha_x(x,t) =&\Sigma^{++}(x)\alpha(x,t) +\Sigma^{+-}(x) \beta(x,t) \nonumber \\
+\int_0^x\mathbf{C}^+(x,\xi)\alpha(\xi,t) d\xi &+ \int_0^x\mathbf{C}^-(x,\xi)\beta(\xi,t)d\xi,
\label{tarpd1}\\
\beta_t\left(x, t\right)-\Lambda^{-} \beta_x(x, t)&=0, \label{tarpd2}\\
\alpha(0,t)  &= {Q} \beta(0, t), \label{tarpd3} \\
\beta(L, t) &= d(t). \label{tarpd4}
\end{align}
The matrix-valued coefficients $ \mathbf{C}^+(x,\xi)\in \mathbb{R}^{3\times 3}$ and $\mathbf{C}^-(x,\xi)\in \mathbb{R}^{3\times1}$ share the same triangular domain $\mathcal{T}$. To design the triggering mechanism, we establish dynamic criteria based on the temporal evolution of $d(t)$. Choosing the Lyapunov functional:
\begin{align}
     V(t) = \int_0^L \sum_{i=1}^3 \frac{A_i}{\lambda_i}\mathrm{e}^{-\frac{\mu x}{\lambda_i}} \alpha_i^2(x,t) + \frac{B}{\Lambda^-} \mathrm{e}^{\frac{\mu x}{\Lambda^-}} \beta^2(x,t) dx,
     \label{lyapunov}
\end{align}
where the constant coefficients $A_1$, $A_2$, $A_3$, $B$ and $\mu$ are positive. The Lyapunov functional is equivalent to the $L_2$ norm of the target system's states, therefore, there exist two constants $p_3>0$ and $p_4>0$ such that
\begin{align}
    p_3\norm{\vartheta}_{L^2}^2 \leq V(t) \leq p_4\norm{\vartheta}_{L^2}^2. \label{eq_norm_V0}
\end{align}
The well-posedness issue of the closed-loop system is essential because the system needs to admit a unique solution under ETC. For the continuous controller $\Bar{U}(t)$ in~\eqref{control_law}, the closed-loop system~\eqref{clpsys1}-\eqref{clpsys4} admits a unique solution with a given condition $(\mathbf{w}^+(\cdot,t_k),\mathbf{w}^-(\cdot,t_k))^\top$ $\in$ $L^2((0,L),\mathbb{R}^4)$~\cite{coron2016stabilization}. Using the method of characteristics and successive approximations in~\cite{hu2015control}, the backstepping kernels of the backstepping transformation are also well-posed. For the closed-loop system under ETC, using the method of characteristics, we can conclude that there exists a unique solution to~\eqref{tarpd1}-\eqref{tarpd4} with a piecewise control input for all $t \geq 0$ in the interval $[t_k,t_{k+1}]$. Then the inverse backstepping transformation~\eqref{invback} leads to the existence of a unique solution to the system \eqref{clpsys1}-\eqref{clpsys4}.
\subsection{Dynamic triggering condition}
This section formulates the event-triggered mechanism (ETM) using a dynamic triggering condition, which can be derived by control discrepancy in~\eqref{dt} and an auxiliary dynamic variable $m(t)$.
\begin{defi}
    Considering the Lyapunov functional candidate $V(t)$ defined in~\eqref{lyapunov}. The event-triggered controller~\eqref{control law sam} with a dynamic event-triggered mechanism, operates at execution time $t_k \geq 0$ (initialized at $t_0 = 0$) forming a finite sequence determined by:
    \begin{itemize}
        \item \text{if} \{$t>t_k$ $\wedge $ $\zeta B \mathrm{e}^{\frac{\mu L}{\Lambda^-}}d^2(t)$ $\geq$ $ \zeta \mu \sigma V(t) - m(t)$\} = $\emptyset$, then the set of the times of the events is \{ $t_0,\dots,t_k$\}.
        \item  \text{if} \{$t>t_k$ $\wedge $ $ \zeta B \mathrm{e}^{\frac{\mu L}{\Lambda^-}}d^2(t)$ $\geq$ $ \zeta \mu \sigma V(t) - m(t)$\} $\ne$ $\emptyset$, then the next execution time is determined by:
        $t_{k+1}$ $=$ $\inf\{ t>t_{k} \wedge  B \mathrm{e}^{\frac{\mu L}{\Lambda^-}}d^2(t)$ $\geq  \zeta \mu \sigma V(t) - m(t)\}$,
    \end{itemize}
    where $m(t)$ satisfies the ordinary differential equation,
    \begin{align}
        \Dot{m}(t) = &- \eta m(t) +  B \mathrm{e}^{\frac{\mu L}{\Lambda^-}}d^2(t) - \sigma \mu V(t) \nonumber\\
        &- \sum_{i=1}^3 \varsigma_i\alpha_i^2(L,t) - \varsigma_4 \beta^2(0,t),
        \label{mdyna}
    \end{align}
    where $\zeta>0$, $\mu>0$, $\sigma > 0$, $\varsigma_i>0$, $i\in \{1,2,3,4\}$, $\eta >0$ and $m(0) = m^0 \in \mathbb{R}^-$.
    \label{defdynamic}
\end{defi}
Combining the event-triggering mechanism and the backstepping controller, we get the algorithm procedure of our proposed method, as shown in Algorithm~\ref{algETC}. It gives the general steps of ETC for the mixed-autonomy traffic system.
\begin{algorithm}[ht]
\caption{The proposed ETC algorithm}\label{algETC}
\KwIn{Model parameters, Designed parameters, Initial conditions}
\KwOut{Traffic density and speed for HVs and AVs}
$t \gets \text{Time step}$\\
$T \gets \text{Ending time}$\\
Formulating the model with corresponding model parameters and designed parameters\\
Solving the backstepping kernels $\mathbf{K}(x,\xi), M(x,\xi)$\\
Using Upwind scheme to discretize the PDE model\\
\While{$ t \leq T $}{
    Calculating the continuous control input $\bar{U}(t)$\\
  \eIf{ Triggering condition is satisfied}{
    Updating the control value to the continuous control input
  }
  {
    Keeping the control value as the same with the last time step
  }
  Applying the control input $\bar{U}_d(t)$ to the system\\
  Updating the traffic states\\
}
\end{algorithm}
Using the dynamic triggering condition in Definition~\ref{defdynamic}, we get an estimate for the dynamic variable $m(t)$.
\begin{lem}
    Under the ETM in Definition \ref{defdynamic}, it holds that $\zeta B \mathrm{e}^{\frac{\mu L}{\Lambda^-}}d^2(t)$ $-$ $ \zeta \mu \sigma V(t) + m(t)$ $\leq 0$ with $m(t) < 0$
    \label{mbound}
\end{lem}
\begin{proof}
    As established in Definition~\ref{defdynamic}, the ETM ensures the system persistently satisfies the following condition:
    \begin{align}
        \zeta B \mathrm{e}^{\frac{\mu L}{\Lambda^-}}d^2(t) -  \zeta \mu \sigma V(t) \leq - m(t).
    \end{align}
    We have the result
    \begin{align}
         B \mathrm{e}^{\frac{\mu L}{\Lambda^-}}d^2(t) -  \mu \sigma V(t) \leq -\frac{1}{\zeta} m(t).
    \end{align}
    Using \eqref{mdyna}, we get 
    \begin{align}
        \Dot{m}(t) \leq -\eta m(t) - \frac{1}{\zeta} m(t) -\sum_{i=1}^3 \varsigma_i \alpha_i^2(L,t) - \varsigma_4\beta^2(0,t).
    \end{align}
    Using the comparison principle, thus 
    \begin{align}
        m(t) < 0 , \forall t \geq 0,
    \end{align}
    this finishes the proof of Lemma \ref{mbound}.
\end{proof}
By taking the derivative of the control discrepancy $d(t)$, we get the estimate of the time derivative of $d(t)$.   {We assume the control discrepancy $d(t)$ is continuous and it serves as the sign of an triggered event and is also used for proving the avoidance of Zeno phenomenon.}
\begin{lem}
    There exists $\epsilon_i >0, i\in\{1,2,3\}$, $\phi_1$ and $\phi_2 > 0$, and $c_4 >0$ for the $d(t)$ introduced in \eqref{dt} with $t \in (t_k, t_{k+1})$, such that
    \begin{align}
        \Dot{d}^2(t) \leq \sum_{i=1}^3\epsilon_i \alpha^2_i(L,t) + \phi_1 d^2(t)+ \phi_2 V(t) +c_4.
    \end{align}
    \label{bounddt}
\end{lem}
\begin{proof}
     {We assume that the system remains well-posed and does not exhibit blow-up under the proposed controller due to the regularity of the traffic system. Taking time derivative of $d(t)$, thus
    \begin{align}
        \Dot{d}(t) =& - \int_0^L \mathbf{L}(L,\xi)\alpha_t(\xi,t) + N(L,\xi) \beta_t(\xi,t)d\xi\nonumber\\
        &+ R\mathbf{w}^+_t(L,t).
    \end{align}
    From the definition of the backstepping transformation (20), we know that $\mathbf{w}^+(x,t)=\alpha(x,t)$, then we using the notation
    \begin{align}
        \dot d(t) =: S(t) + R\alpha_t(L,t).\label{defdt}
    \end{align}
    where $S(t)=:- \int_0^L \big(L(L,\xi)\,\alpha_t(\xi,t)+N(L,\xi)\,\beta_t(\xi,t)\big)\,d\xi$.
    Using the dynamics of target system at $x=L$, we have
    \begin{align}
    \alpha_t(L,t) = -\Lambda^+ \alpha_x(L,t) + G(L,t),
    \end{align}
    with
    \begin{align}
    G(L,t) =:& \Sigma^{++}(L)\alpha(L,t) + \Sigma^{+-}(L)\,d(t) \nonumber\\
    &+ \int_0^L \!\big(\mathbf{C}^+(L,\xi)\alpha(\xi,t)+\mathbf{C}^-(L,\xi)\beta(\xi,t)\big)\,d\xi. \label{defG} 
    \end{align}
    Then we have
    \begin{align}
        R\,\alpha_t(L,t) = -R\Lambda^+\alpha_x(L,t) + R\,G(L,t).
    \end{align}
    And we get the upper bound of the square of $R\,\alpha_t(L,t)$ as
    \begin{align}
        \norm{R \alpha_t(L,t)}^2 \leq 2 \norm{R \Lambda^+}^2 \norm{\alpha_x(L,t)}^2 + 2 \norm{R}^2 \norm{G(L,t)}^2
    \end{align}
    Next, we will give the bound of $S(t)$ and $G(L,t)$.
    }
    
      {
    We already know the dynamics of the system~\eqref{tarpd1}-\eqref{tarpd4} and then integrating by parts, the expression of $S(t)$ is obtained as
    \begin{align}
        &S(t) = \mathbf{L}(L,L)\Lambda^+\alpha(L,t) - N(L,L)\Lambda^- \beta(L,t) \nonumber\\
        &+ \left( N(L,0)\Lambda^- - \mathbf{L}(L,0)\Lambda^+Q \right)\beta(0,t)\nonumber\\
        &- \int_0^L \mathbf{L}_{\xi}(L,\xi)\Lambda^+\alpha(\xi,t)d\xi + \int_0^L N_{\xi}(L,\xi)\Lambda^-\beta(\xi,t)d\xi \nonumber\\
        &-\int_0^L \mathbf{L}(L,\xi)\Sigma^{++}(\xi)\mathbf{w}^+(\xi,t)d\xi \nonumber\\
        &- \int_0^L \mathbf{L}(L,\xi)\Sigma^{+-}(\xi)\mathbf{w}^-(\xi,t)d\xi.
    \end{align}
    Using Young's inequality for the square of $S(t)$, and then using Cauchy-Schwarz inequality, the estimate of the $S(t)$ is obtained as: 
    \begin{align}
       \norm{S(t)}^2 \leq& 8\sum_{i=1}^3 \ell_i^2(L,L)\lambda_i^2\alpha_i^2(L,t) + 8 N^2(L,L)(\Lambda^-)^2 d^2(t)\nonumber\\
       &+ \frac{8}{p_3} \left(c_1 + \frac{c_2}{p_1}\right)V(t),
    \end{align}
    where 
    $c_1 $ $=$ $ \max\{\int_0^L$ $\norm{\mathbf{L}_\xi(L,\xi)\Lambda^+}^2 d\xi,$ $ \int_0^L$ $\norm{N_\xi(L,\xi)\Lambda^-}^2 d\xi \}$
    $c_2 $ $ =$ $ \max\{\int_0^L \norm{\mathbf{L}(L,\xi) \Sigma^{++}(\xi)}^2 d\xi,$ $ \int_0^L \norm{\mathbf{L}(L,\xi) \Sigma^{+-}(\xi)}^2 d\xi \}$.
    Next we will bound the term $G(L,t)$, recalling the definition of $G(L,t)$ in~\eqref{defG},
    \begin{align}
        G(L,t) = & \Sigma^{++}(L)\alpha(L,t) +  \Sigma^{+-}(L)\,d(t) \nonumber\\
    &+ \int_0^L \! \big(\mathbf{C}^+(L,\xi)\alpha(\xi,t)+\mathbf{C}^-(L,\xi)\beta(\xi,t)\big)\,d\xi.
    \end{align}
    Using Young's inequality and Cauchy-Schwarz inequality again for square of $G(L,t)$, we obtain
    \begin{align}
        \norm{G(L,t)}^2 \leq& 4 \sum_{i=1}^3 j_i^2 \alpha_i^2(L,t) + 4(R \Sigma^{+-}(L))^2 \dot d^2(t) \nonumber\\
        &+ \frac{4c_3}{p_3} V(t),
    \end{align}
    where $c_3 = \max \{\int_0^L\norm{ \mathbf{C}^+(L,\xi)}^2 d\xi,\int_0^L\norm{ \mathbf{C}^-(L,\xi)}^2 d\xi\} $, $j_i$ denotes the element in coefficient matrix $\Sigma^{++}(x)$. Taking the bound of $S(t)$ and $G(L,t)$ into~\eqref{defdt} , we finally get the estimate of $\dot d^2(t)$ as 
    \begin{align}
        \dot d^2(t) \leq \sum_{i=1}^3\epsilon_i \alpha^2_i(L,t) + \phi_1 d^2(t)+ \phi_2 V(t) + c_4,
    \end{align}
    where
    \begin{align*}
        \epsilon_i &= (8 \ell_i^2(L,L)\lambda_i^2 + 8 j_i \norm{R}^2),\\
        \phi_1 &= (8 N^2(L,L)(\Lambda^-)^2 + 2(R \Sigma^{+-}(L))^2)\norm{R}^2,\\
        \phi_2 &=  \frac{8}{p_3} \left(c_1 + \frac{c_2}{p_1}\right)+\frac{8c_3\norm{R}^2}{p_3}, \\
        c_4 &\geq 2 \norm{R \Lambda^+}^2 \norm{\alpha_x(L,t)}^2,
    \end{align*}
    where $c_4$ is the upper bound of term $2 \norm{R \Lambda^+}^2 \norm{\alpha_x(L,t)}^2$.
     This concludes the proof of Lemma \ref{bounddt}.
     }
\end{proof}

Both Lemma~\ref{mbound} and Lemma~\ref{bounddt} serve as an intermediate bridge to show the avoidance of the Zeno phenomenon of the system. After the two estimates are obtained, we could define another function to prove that there exists a minimal dwell-time between two triggering times in our ETC framework.

\subsection{Avoidance of Zeno phenomenon}
The Zeno phenomenon should always be avoided in our control design. We have the following theorem to guarantee the Zeno free for the closed-loop system~\eqref{clpsys1}-\eqref{clpsys4}.
\begin{thm}
    The dynamic triggering mechanism in Definition~\ref{defdynamic} guarantees a minimum dwell-time $\tau^\star>0$ between two consecutive triggering times, satisfying $t_{k+1} - t_k \geq \tau^\star, k \geq 0$. The parameters $\zeta$, $\mu$, $\sigma$, $\varsigma_i,i \in\{1,2,3,4\}$, $\eta$, $\epsilon_i, i \in \{1,2,3\}$ satisfy:
    \begin{align}
        \varsigma_i &\geq \max\{\zeta B \mathrm{e}^{\frac{\mu L}{\Lambda^-}}\epsilon_i,\zeta\mu\epsilon_i, i \in \{1,2,3\}\},\label{select1}\\
        \varsigma_4 &\geq \max\{0, -2\zeta\mu(\sum A_iq_i^2 - B), i \in \{1,2,3\}\}.\label{select2}
    \end{align}
    \label{minimalthm}
\end{thm}
\begin{proof}
    Based on Definition~\ref{defdynamic}, for all time $t \geq 0$, all executed events satisfy:
    \begin{align}
        \zeta B \mathrm{e}^{\frac{\mu L}{\Lambda^-}}d^2(t) \leq  \zeta \mu \sigma V(t) - m(t).
    \end{align}
    Defining another function $\Psi(t)$ as
    \begin{align}
        \Psi(t) = \frac{\zeta B \mathrm{e}^{\frac{\mu L}{\Lambda^-}}d^2(t) + \frac{1}{2}m(t)}{ \zeta \mu \sigma V - \frac{1}{2}m(t)}. \label{defPsi}
    \end{align}
    The function $d(t)$ and $V(t)$ are continuous on time interval $[t_{k},t_{k+1}]$, so that the function $\Psi(t)$ is also a continuous function at $[t_{k},t_{k+1}]$. We can derive that there exists $t'_k > t_k$ such that $\forall t \in [t'_{k},t_{k+1}]$, $\Psi(t) \in [0,1]$ using the intermediate value theorem and the term $\zeta \mu \sigma V - \frac{1}{2}m(t)$ is inherently avoided to be zero. The time derivative of function $\Psi(t)$ is then obtained as:
    \begin{align}
        \Dot{\Psi}(t) = \frac{2\zeta B \mathrm{e}^{\frac{\mu L}{\Lambda^-}} d \Dot{d} + \frac{1}{2}\Dot{m}}{ \zeta \mu \sigma V - \frac{1}{2}m} - \frac{\zeta  \mu \sigma \Dot{V} - \frac{1}{2}\Dot{m}}{ \zeta \mu \sigma V - \frac{1}{2}m}\Psi.
    \end{align}
    Using Young's inequality, we have
    \begin{align}
         &\Dot{\Psi}(t) \leq \frac{\zeta B \mathrm{e}^{\frac{\mu L}{\Lambda^-}}d^2}{ \zeta \mu \sigma V - \frac{1}{2}m} + \frac{\zeta B \mathrm{e}^{\frac{\mu L}{\Lambda^-}}\Dot{d}^2}{ \zeta \mu \sigma V - \frac{1}{2}m} \nonumber\\
         & +\frac{\frac{1}{2}(-\eta m +B \mathrm{e}^{\frac{\mu L}{\Lambda^-}}d^2 - \sigma\mu V)}{\zeta \sigma \mu V - \frac{1}{2}m}\nonumber\\
         & + \frac{\frac{1}{2}\left( - \sum_{i=1}^3 \varsigma_i\alpha_i^2(L,t) - \varsigma_4 \beta^2(0,t)\right)}{ \zeta \mu \sigma V - \frac{1}{2}m} -\frac{\zeta \mu \Dot{V}\Psi}{ \zeta \mu \sigma V - \frac{1}{2}m} \nonumber\\
         & + \frac{\frac{1}{2}\left( -\eta m +B \mathrm{e}^{\frac{\mu L}{\Lambda^-}}d^2 - \sigma\mu V \right)}{ \zeta \mu \sigma V - \frac{1}{2}m}\Psi \nonumber\\
         & +\frac{\frac{1}{2}\left( - \sum_{i=1}^3 \varsigma_i\alpha_i^2(L,t) - \varsigma_4 \beta^2(0,t)\right)}{ \zeta \mu \sigma V - \frac{1}{2}m} \Psi.
    \end{align}
    Based on the Lyapunov functional definition in~\eqref{lyapunov}, the time derivative $\Dot{V}(t)$ is derived via integration by parts and combined with boundary conditions of the perturbed target system. This yields the expression for $\Dot{V}(t)$:
    \begin{align}
        \Dot{V} &\leq - \sum_{i=1}^3 A_i \mathrm{e}^{-\frac{\mu L}{\lambda_i}}\alpha_i^2(L,t) +  (\sum_{i=1}^3 A_i q_i^2 - B) \beta^2(0,t) \nonumber\\
        & + B\mathrm{e}^{\frac{\mu L}{\Lambda^-}}d^2(t)-(\mu-\gamma) V,
    \end{align}
    where $\gamma = \frac{2A}{p_3\min\{ \lambda_i\}}( \max_{x\in [0,L]} \norm{\Sigma^{++}(x)} $ $+ (1+\frac{1}{p_1})$ $ \max_{x\in [0,L]} \norm{\Sigma^{+-}(x)})$.
    Replacing $\Dot{V}$ and using the result in Lemma~\ref{bounddt}, we have
        \begin{align}
        &\Dot{\Psi}(t) \leq - \frac{(\zeta\mu(\sum A_iq_i^2 - B)+\frac{1}{2}\varsigma_4)\beta^2(0,t)}{\zeta \mu \sigma V - \frac{1}{2}m}\Psi \nonumber\\
        &- \frac{\frac{1}{2}\eta m}{\zeta \mu \sigma V - \frac{1}{2}m}\Psi + \frac{\sum (\zeta \mu \epsilon_i - \varsigma_i)\alpha_i^2(L,t)}{\zeta \mu \sigma V - \frac{1}{2}m}\Psi \nonumber\\
        & + \frac{B \mathrm{e}^{\frac{\mu L}{\Lambda^-}}d^2(-\zeta \mu \sigma +\frac{1}{2})}{\zeta \mu \sigma V - \frac{1}{2}m}\Psi + \frac{(\zeta \mu \sigma(\mu - \gamma)-\frac{1}{2}\mu \sigma)V}{\zeta \mu \sigma V - \frac{1}{2}m}\Psi \nonumber\\
        &+\frac{\zeta B \mathrm{e}^{\frac{\mu L}{\Lambda^-}}(1 + \phi_1+\frac{1}{2\zeta}) d^2}{\zeta \mu \sigma V - \frac{1}{2}m} + \frac{\sum(\zeta B\mathrm{e}^{\frac{\mu L}{\Lambda^-}} \epsilon_i  - \varsigma_i) \alpha_i^2(L,t) }{\zeta \mu \sigma V - \frac{1}{2}m}\nonumber\\
        & + \frac{(\zeta B \mathrm{e}^{\frac{\mu L}{\Lambda^-}}\phi_2 - \frac{1}{2} \mu \sigma)V}{\zeta \mu \sigma V - \frac{1}{2}m} - \frac{\frac{1}{2}\eta m}{\zeta \mu \sigma V - \frac{1}{2}m} - \frac{\frac{1}{2}\varsigma_4\beta^2(0,t)}{\zeta \mu \sigma V - \frac{1}{2}m}.
    \end{align}
    Choosing $\varsigma_i \geq \zeta B \mathrm{e}^{\frac{\mu L}{\Lambda^-}}\epsilon_i$ and $\varsigma_i \geq \zeta\mu\epsilon_i$, $i\in \{1,2,3\}$, $\varsigma_4 > 0$ and $\varsigma_4 +2\zeta\mu(\sum A_iq_i^2 - B) >0$, we get the following equation after simplification
    \begin{align}
        &\Dot{\Psi}(t) \leq \frac{\zeta B \mathrm{e}^{\frac{\mu L}{\Lambda^-}}(1 + \phi_1+\frac{1}{2\zeta}) d^2}{\zeta \mu \sigma V - \frac{1}{2}m}  + \frac{(\zeta B \mathrm{e}^{\frac{\mu L}{\Lambda^-}}\phi_2 - \frac{1}{2} \mu \sigma)}{\zeta \mu \sigma}+\eta\nonumber \\
        &+ \frac{B \mathrm{e}^{\frac{\mu L}{\Lambda^-}}d^2(-\zeta \mu \sigma+\frac{1}{2})}{\zeta \mu \sigma V - \frac{1}{2}m}\Psi +\eta \Psi  + \frac{(\zeta \mu \sigma (\mu - \gamma)-\frac{1}{2}\mu \sigma)}{\zeta \mu \sigma }\Psi.
    \end{align}
    Rewriting and rearranging the above equation, thus the we deduce that
    \begin{align}
        \Dot{\Psi}(t) \leq&  \left(\frac{-\zeta\mu + \frac{1}{2}}{\zeta}\right)\Psi^2 +\left( 1+\phi_1 + \frac{1}{2\zeta} \right.\nonumber\\
        &+ \left.\frac{-\zeta\mu \sigma + \frac{1}{2}}{\zeta} + \eta + \frac{\zeta\mu \sigma(\mu-\gamma) - \frac{1}{2}\mu \sigma}{\zeta\mu \sigma} \right)\Psi\nonumber\\
        &+\left( \frac{(\zeta B \mathrm{e}^{\frac{\mu L}{\Lambda^-}}(\phi_2+c_4) - \frac{1}{2} \mu \sigma)}{\zeta \mu \sigma} + \eta + 1+\phi_1+\frac{1}{2\zeta}\right).
    \end{align}
    And the $\Psi(t)$ has the form
    \begin{align}
        \dot \Psi(t) \leq \varphi_1\Psi^2(t) + \varphi_2 \Psi(t) + \varphi_3,
    \end{align}
    where 
    $\varphi_1 = \frac{1}{2\zeta} - \mu \sigma$,
    $  {\varphi_2} = 1+ \phi_1 + \frac{1}{2\zeta} (1-\sigma)\mu -\gamma +\eta$,
      {$\varphi_3 = \frac{B \mathrm{e}^{\frac{\mu L}{\Lambda^-}}(\phi_2+c_4)}{\mu \sigma} + 1+\eta + \phi_1$}. 
    Using the comparison principle, we get the time from $\Psi(t_k') = 0$ to $\Psi(t_{k+1}) = 1$ is at least
    \begin{align}
        \tau^\star = \int_0^L \frac{1}{\varphi_1 s^2 + \varphi_2s +\varphi_3} ds.
    \end{align}
    Then, $t_{k+1} - t_{k} \geq t_{k+1} - t_{t'_k} = \tau^\star$. This concludes the proof of Theorem \ref{minimalthm}.
\end{proof}
Building upon the established minimal dwell-time guarantee between consecutive triggering instants, Zeno behavior is eliminated. This result, combined with the preceding analysis, rigorously establishes exponential stability for the closed-loop dynamics \eqref{clpsys1}-\eqref{clpsys4} under the event-triggered controller \eqref{control law sam}, as formalized in Theorem~\ref{esthm}.
\begin{thm}
    Let $A_i >0, i\in\{1,2,3\}$, $B>0$, $\zeta>0$, $\eta \in (0,1)$, $\varsigma_i,i\in\{1,2,3,4\} \in (0,1)$ such that
    \begin{align}
        \varsigma_i - A_i \mathrm{e}^{-\frac{\mu L}{\lambda_i}} &\leq 0 , i \in \{1,2,3\},\label{select3}\\
        \varsigma_4 + \sum_{i=1}^3 A_i q_i^2 - B &\leq 0, i \in \{1,2,3\},\label{select4}
    \end{align}
    The system~\eqref{clpsys1}-\eqref{clpsys4} with the ETC~\eqref{control law sam}   {achieves exponential convergence in $L^2$-sense}.
    \label{esthm}
\end{thm}
\begin{proof}
    Consider the following Lyapunov functional for system~\eqref{tarpd1}-\eqref{tarpd4},
    \begin{align}
        V_{d}(t,m) = V(t) - m(t).
    \end{align}
    The time derivative of the Lyapunov functional is:
    \begin{align}
        \Dot{V}_{d}(t,m) &\leq  B\mathrm{e}^{\frac{\mu L}{\Lambda^-}}d^2(t)-(\mu-\gamma) V - \Dot{m}(t)\nonumber\\
        & - \sum_{i=1}^3 A_i \mathrm{e}^{-\frac{\mu L}{\lambda_i}}\alpha_i^2(L,t) +  (\sum_{i=1}^3 A_i q_i^2 - B) \beta^2(0,t),    
    \end{align}
    we already define $\Dot{m}(t)$ in Definition~\ref{defdynamic}, thus
    \begin{align}
        \Dot{V}_{d} \leq &-(\mu - \gamma) V + B\mathrm{e}^{\frac{\mu L}{\Lambda^-}}d^2(t) + \eta m + \mu \sigma V \nonumber\\
        &-B \mathrm{e}^{\frac{\mu L}{\Lambda^-}}d^2(t) + \sum_{i=1}^3 (\varsigma_i - A_i \mathrm{e}^{-\frac{\mu L}{\lambda_i}})\alpha_i^2(L,t) \nonumber\\
        &+ (\varsigma_4 + \sum_{i=1}^3 A_i q_i^2 - B) \beta^2(0,t).
    \end{align}
    Simplifying the above equation, it can be deduced that
    \begin{align}
        \Dot{V}_{d} \leq -(\mu(1-\sigma) - \gamma){V}_{d} + (\eta - (\mu(1-\sigma) - \gamma))m.
    \end{align}
    Choosing $\eta - (\mu(1-\sigma) - \gamma) \geq 0$, we get
    \begin{align}
         \Dot{V}_{d} \leq -(\mu(1-\sigma) - \gamma){V}_{d}.
    \end{align}
      {Using the fact $V(t) \leq V_d(t,m)$ from the definition $V_{d}(t,m) = V(t) - m(t)$ and the property $m(t) \leq 0$, we get 
    \begin{align}
        V(t) \leq \mathrm{e}^{(-(\mu(1-\sigma) - \gamma)t)}(V(0)- m(0)).
    \end{align}
    Therefore, an estimation of the original system of the $L_2$ norm can be written as
    \begin{align}
        \norm{\mathbf{w}(x,t)}^2_{L^2} \leq \frac{p_2p_4}{p_1 p_3} \mathrm{e}^{-(\mu(1-\sigma) - \gamma)t}\left(\norm{\mathbf{w}(x,0)}^2_{L^2} - m(0)\right).\label{estimate4ori}
    \end{align}}
    This concludes the proof of Theorem~\ref{esthm}.
\end{proof}
  {
\begin{remark}
    In Theorem~\ref{esthm}, we have established the exponential convergence for the close-loop mixed-autonomy traffic system to the equilibrium point. The exponential stability for the mixed-autonomy traffic system could be obtained by taking $m(0) = 0$. However, we assume the initial value of $m$ is negative and the function $\Psi(t)$ may be undefined in~\eqref{defPsi} when $m(0)=0$ and the result $\Psi(t)\in [0,1]$ may not be obtained through intermediate value theorem. As a result, the minimal dwell time between two triggered events could not be proved through our analysis in the proof of Theorem~\ref{minimalthm}.
\end{remark}
}
  {
\begin{remark}
    The designed parameters $\zeta,\eta,\mu,B,\zeta$, $\varsigma_i, i \in \{1,2,3,4\}$, $A_i, i \in \{1,2,3\}$ are selected by condition~\eqref{select1}-\eqref{select2} in Theorem~\ref{minimalthm} and~\eqref{select3}-\eqref{select4} in Theorem~\ref{esthm}. The parameter $\sigma$, $\mu$ are related to the decay rate of the Lyapunov function. However, the optimal choice of design parameters regarding to the conservatism and sampling speed is not tackled, which could be further explored in the future work. In this paper, we mainly focus on the stability result and the traffic application of the proposed ETC. A sensitive analysis of design parameters was provided in the next section.
\end{remark}
}

\subsection{Observer-based event-triggered controller}


  {In previous section, we developed the ETC design using full-state information which can be difficulty to obtain in practice.  Only sparse measurement of traffic speed and density is collected by a  limited number of loop detectors and roadside cameras. Therefore,  it is necessary to design boundary observer for state estimation. Next, we will formulate an observer-based ETC for the mixed-autonomy traffic system.}

  {First, consider an anti-collocated boundary observer that relies on the measurement $y(t) = \mathbf{w}^-(0,t)$, the observer equations are given as:
\begin{align}
        \mathbf{\hat{w}}^+_t(x,t) &+\Lambda^{+} \mathbf{\hat{w}}^+_x(x,t) = \Sigma^{++}(x)\mathbf{\hat{w}}^+ (x,t)\nonumber\\
        &+\Sigma^{+-}(x) \mathbf{\hat{w}}^-(x, t) + {P}^+(x)\mathbf{\tilde{w}}^-(0, t),\label{ob-sys1} \\
        \mathbf{\hat{w}}^-_t(x, t)&-\Lambda^{-} \mathbf{\hat{w}}^-_x(x, t) = \Sigma^{-+}(x) \mathbf{\hat{w}}^+(x,t) \nonumber\\
        &+{P}^-(x)\mathbf{\tilde{w}}^-(0, t), \label{ob-sys2}
\end{align}
with boundary conditions,
\begin{align}
    \mathbf{\hat{w}}^+(0,t)  &= {Q} \mathbf{\hat{w}}^-(0, t), \label{ob-sys1-bd} \\
    \mathbf{\hat{w}}^-(L, t) &= {R}\mathbf{\hat{w}}^+(L, t)+\bar{U}_{of}(t), \label{ob-sys2-bd}
\end{align}
where gains of the output injections ${P}^+(x), {P}^-(x)$ need to be designed. 
At the same time, using the backstepping transformation in~\eqref{back4} for the hat-system~\eqref{ob-sys1}-\eqref{ob-sys2-bd}, we get the target system with states $\mathcal{K}\mathbf{w^+}:=$ $\hat{\alpha} = [\hat{\alpha}_1,\hat{\alpha}_2,\hat{\alpha}_3]^\mathsf{T}$, $\mathcal{K}\mathbf{w^-}:=$$\hat{\beta}$. More details of the backstepping transformation, the kernel equations as well as the output injections can be found in~\cite{hu2015control,burkhardt2021stop,espitia2020observer}. 
The output-feedback control law is
\begin{align}
    \bar{U}_{of}(t)=&\int_0^L (\mathbf{L}(L, \xi)\hat{\alpha}(\xi,t) +N(L, \xi) \hat{\beta} (\xi, t)) {d} \xi\nonumber\\
    &-R\hat{\mathbf{w}}^+(L,t),\label{bs_control_of}
\end{align}
and its sampled version is given as
\begin{align}
    \bar{U}_{of}^d(t)=&\int_0^L (\mathbf{L}(L, \xi)\hat{\alpha}(\xi,t_k) +N(L, \xi) \hat{\beta} (\xi, t_k)) {d} \xi \nonumber\\
    &-R\hat{\mathbf{w}}^+(L,t_k).\label{bs_control_of_etc}
\end{align}
The actuation discrepancy $d_o(t)$ of the output-feedback controler is 
\begin{align}
    d_0(t) = &-R(\mathbf{w}^+(L,t_k) - \mathbf{w}^+(L,t)) \nonumber\\ 
    &+ \int_0^L \bigg(\mathbf{L}(L, \xi)(\hat{\alpha}(\xi,t_k)-\hat{\alpha}(\xi,t))\nonumber\\
    &+N(L, \xi) (\hat{\beta} (\xi, t_k)- \hat{\beta} (\xi, t)\bigg) {d} \xi. \label{dtof}
\end{align}
}
  {
\begin{remark}
   We have formulated an observer-based ETC scheme without giving a theoretical proof of convergence for the closed-loop system to rule out Zeno behavior and to establish finite-time convergence of the closed-loop system. This paper mainly focuses on the application relevance of ETC in mixed-autonomy traffic. To illustrate its performance, we present simulation results for the observer-based ETC for the mixed-autonomy traffic system in the experiments section. While the general steps in proof  may follow approaches in~\cite{espitia2020observer,espitia2022traffic,zhao_event-triggered_2026}, extending them to mixed-autonomy traffic systems introduces significant theoretical complexity. In particular, deriving sufficient conditions for stability—and especially for avoiding the Zeno phenomenon—would involve numerous system and design parameters. These conditions would likely need to be expressed in terms of linear matrix inequalities (LMIs), potentially resulting in a highly complex and less interpretable formulation. Given the application focus of this paper, a full mathematical derivation is beyond the scope of the current paper. We regard this as an important theoretical direction for future work.
\end{remark}
}

\section{Experiments}\label{sec5}
In this section, we analyze the performance of the proposed ETC for the mixed-autonomy traffic system. Extensive simulations are performed to test the proposed ETC under different scenarios. 
\subsection{Simulation settings}
\textbf{Model parameters selection}. We run the simulation on a $6.5\text{m}$ wide, $L = 1 \text{km}$ long roadway over $450\text{s}$. The width of the road is comparable to three lanes and the capacity of the road is higher than that of a conventional road due to the lane-free modeling setting and the frequent ``creeping effect," where faster vehicles overtake slower ones. The experiments employ the following model parameters. Equilibrium densities of HVs and AVs are selected as $\rho_{\rm h}^\star = 110\text{veh/km}$, $\rho_{\rm a}^\star = 95 \text{veh/km}$. Equilibrium speeds are $v_{\rm h}^\star = 32\text{km/h}$, $v_{\rm a}^\star = 15\text{km/h}$. The relaxation time are $\tau_{\rm h} = 30\text{s}$, $\tau_{\rm a} = 60 \text{s}$. The pressure exponent value are selected as $\gamma_{\rm h} = 2.5$, $\gamma_{\rm a} = 2$. The spacing polices are $s_{\rm h} = 8 \text{m}$, $s_{\rm a} = 15 \text{m}$. The maximum area occupancy $\overline{AO}_{\rm h} = 0.9$, $\overline{AO}_{\rm a} = 0.85$. The free-flow speed of HVs and AVs are $V_{\rm h} = 80\text{km/h}$ and $V_{\rm a} = 60\text{km/h}$.  To induce stop-and-go oscillations, sinusoidal perturbations are imposed as initial conditions.

\textbf{Design parameters selection}. In addition to the model parameters, the design parameters for Lyapunov analysis and ETC design are also essential. We choose $\zeta = 8\times 10^{-3}$, $\sigma = 1\times10^{-4}$, $\eta = 0.9$, $A_1 = 2\times 10^{-2}$,  $A_2 = 3\times 10^{-3}$,  $A_3 = 4\times 10^{-3}$, $B = 9\times 10^{-3}$, and the other parameters are selected as $\varsigma_1 = 2\times 10^{-10}$, $\varsigma_2 = 2\times 10^{-9}$, $\varsigma_3 = 1.2\times 10^{-12}$, $\varsigma_4 = 0.01$, $\mu = 5\times 10^{-4}$. 

The simulations are performed on an Intel core I9-12900K CPU with a clock rate of 3.6 GHz, and a Nvidia GeForce RTX 4090 Ti GPU.

\subsection{Open-loop and closed-loop results}
\begin{figure}[!tbp]
    \centering
    \subfloat[  {The open-loop density and speed evolution of HVs}]{ \includegraphics[width =0.95\linewidth]{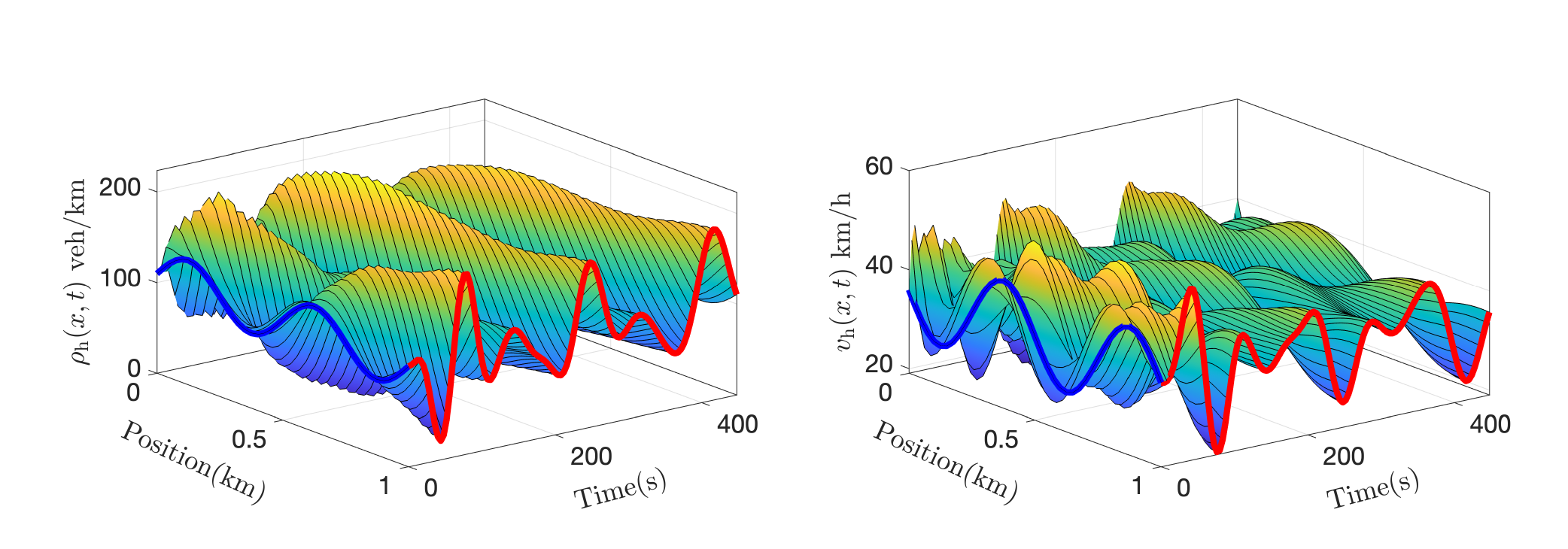}} \\
    \subfloat[  {The open-loop density and speed evolution of AVs}]{\includegraphics[width =0.95\linewidth]{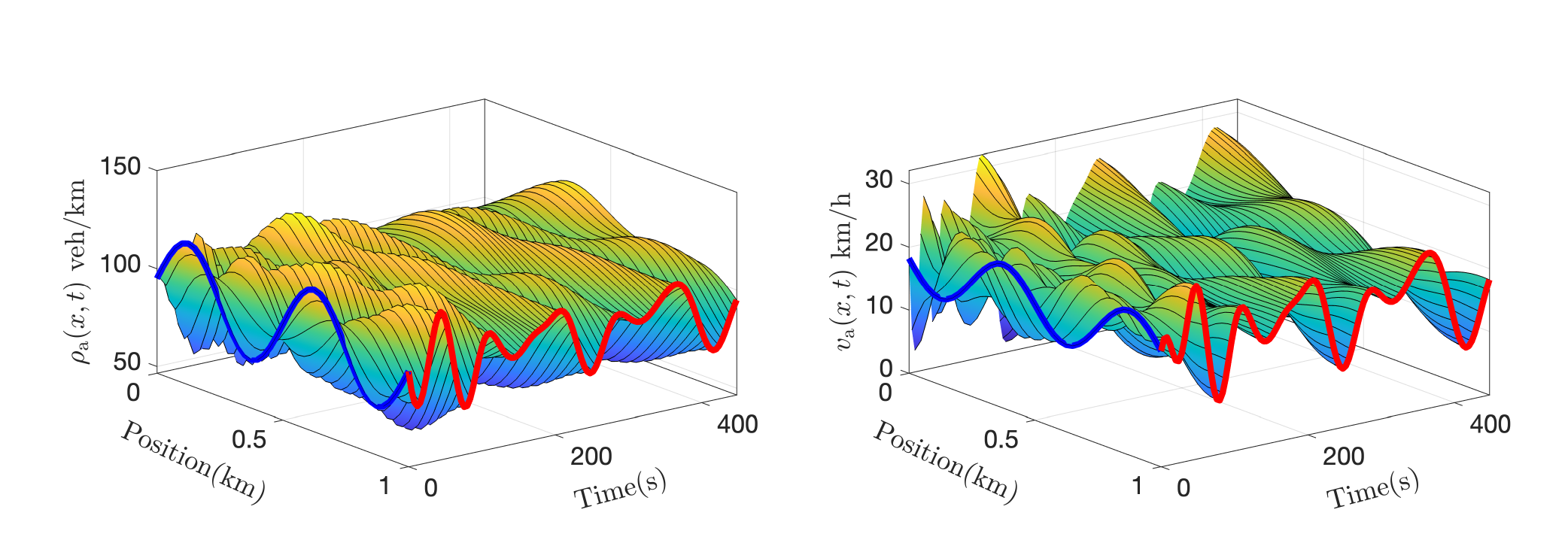}}
    \caption{  {The open-loop traffic density and speed of mixed-autonomy traffic system. Traffic states oscillate in the simulation period.}}
    \label{open-loop}
\end{figure}

\begin{figure}[!tbp]
    \centering
    \subfloat[  {The closed-loop density and speed evolution of HVs}]{\includegraphics[width =0.95\linewidth]{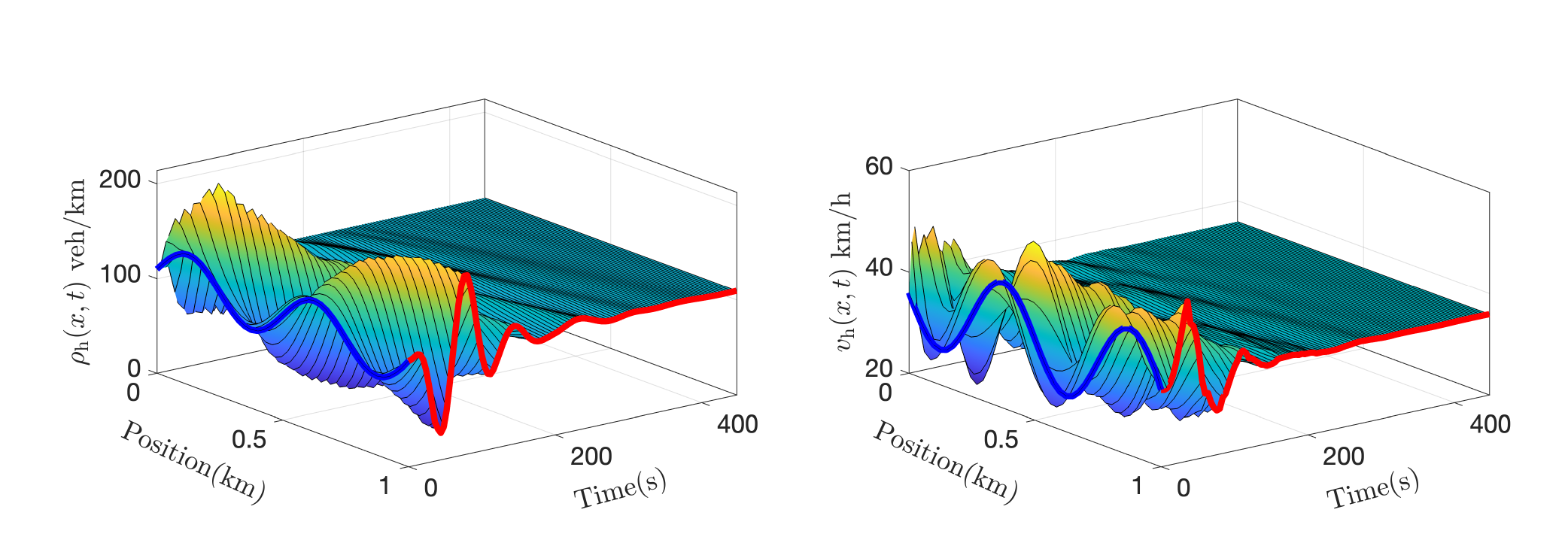}} \\
    \subfloat[  {The closed-loop density and speed evolution of AVs}]{\includegraphics[width =0.95\linewidth]{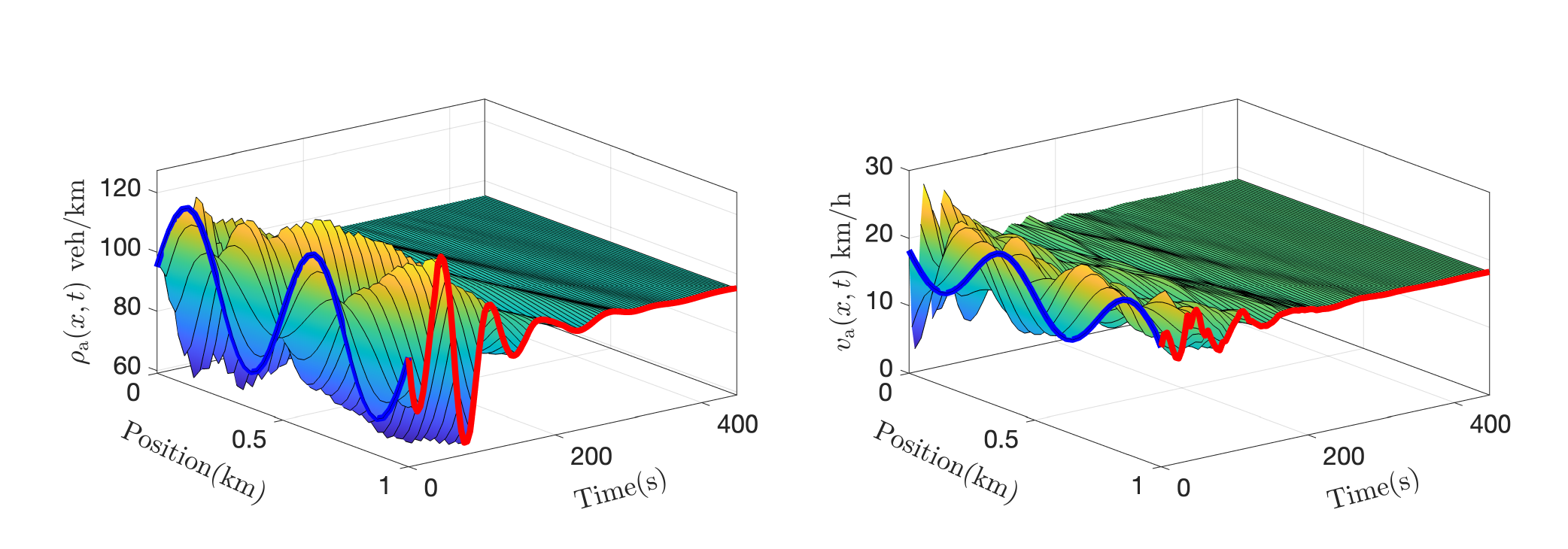}}
    \caption{  {The closed-loop traffic density and speed of mixed-autonomy system under proposed full-state feedback ETC with AV's spacing policy $s_{\rm a}=15\text{m}$}}
    \label{cl_event}
\end{figure}

\begin{figure}[!tbp]
    \centering
    \subfloat[Control input ]{\includegraphics[width= 0.45\linewidth]{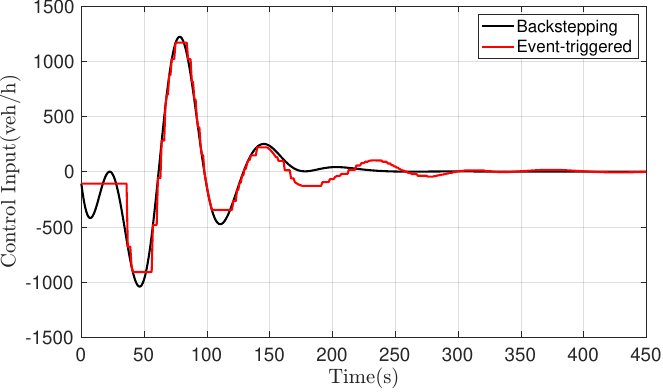}}
    \subfloat[Release instants ]{\includegraphics[width= 0.45\linewidth]{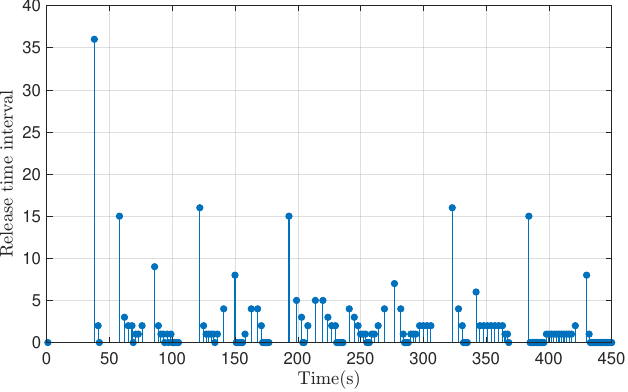}}
    \caption{The performance of ETC under $s_{\rm a}=15\text{m}$}
    \label{event-control}
\end{figure}

\begin{figure}[!tbp]
    \centering
    \subfloat[  {The closed-loop density and speed evolution of HVs}]{\includegraphics[width =0.95\linewidth]{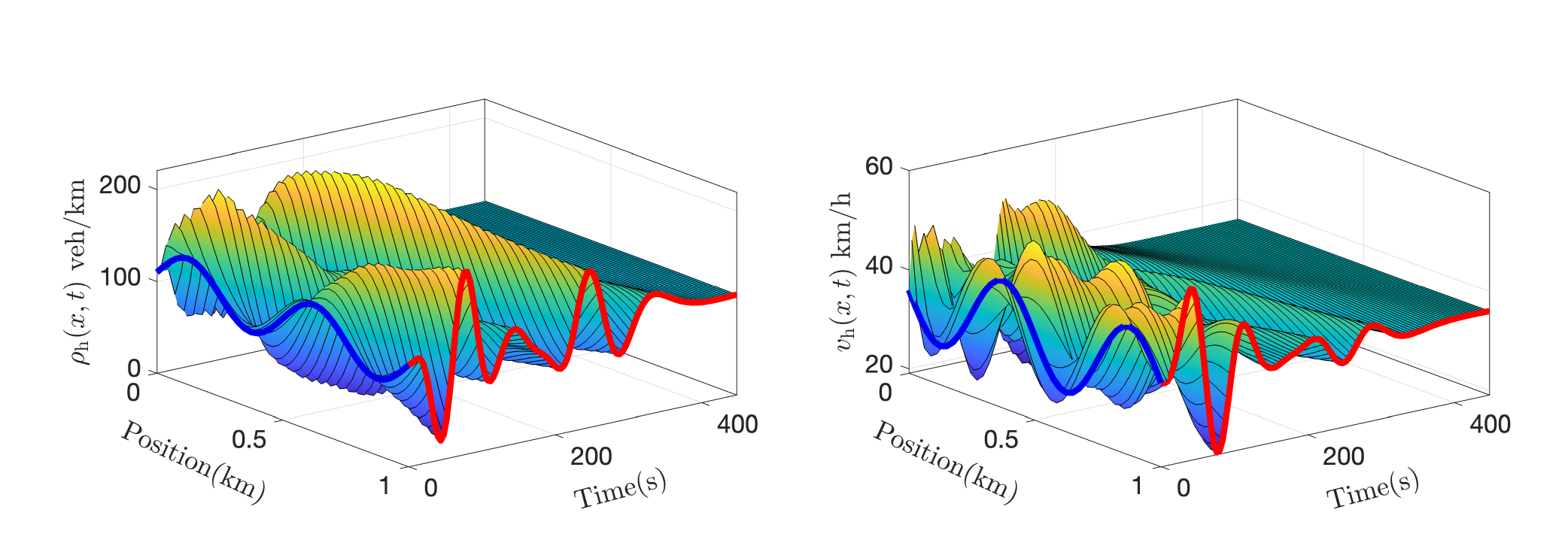}} \\
    \subfloat[  {The closed-loop density and speed evolution of AVs}]{\includegraphics[width =0.95\linewidth]{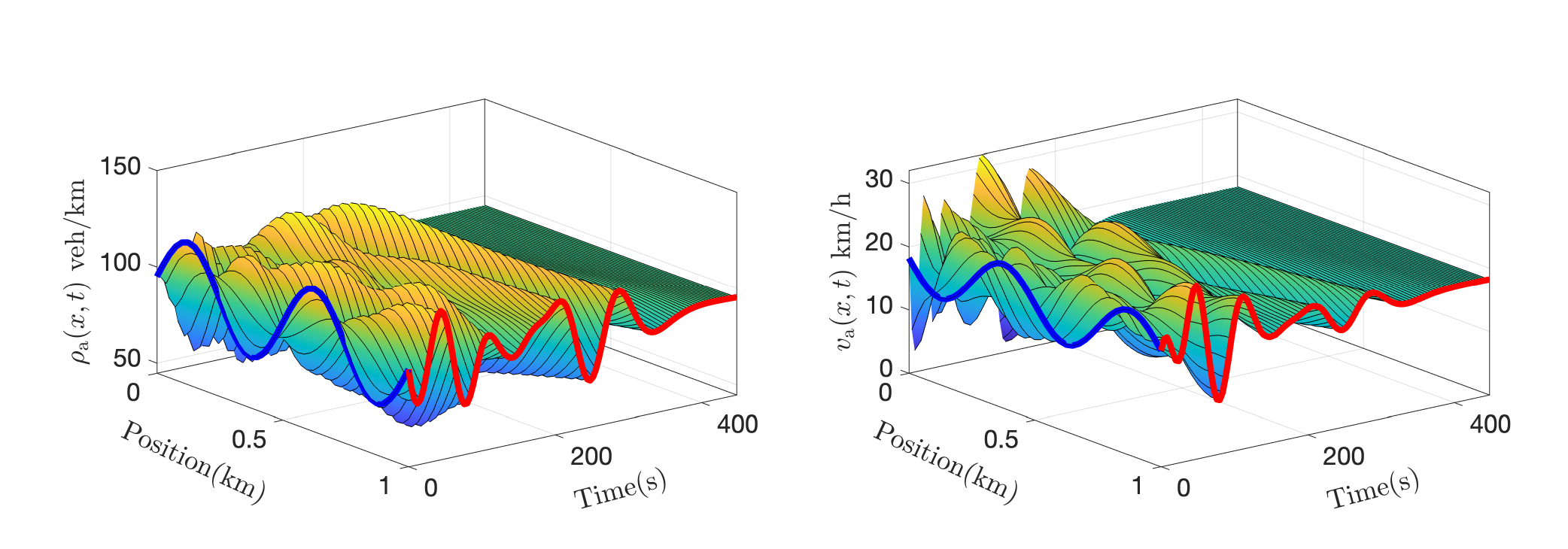}}
    \caption{  {The closed-loop traffic density and speed of mixed-autonomy system under observer-based ETC with AV's spacing policy $s_{\rm a}=15\text{m}$}}
    \label{cl_event_obs}
\end{figure}

\begin{figure}[!tbp]
    \centering
    \subfloat[Control input ]{\includegraphics[width= 0.45\linewidth]{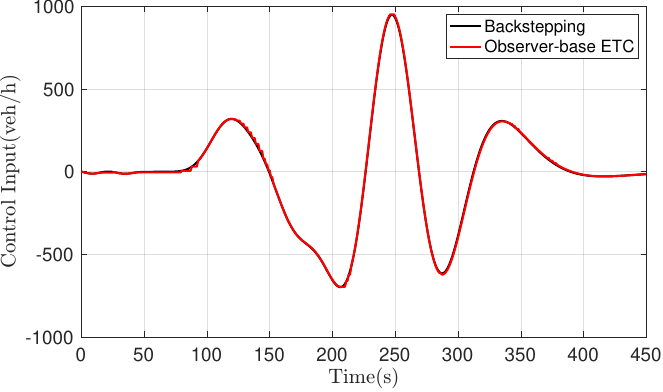}}
    \subfloat[Release instants ]{\includegraphics[width= 0.45\linewidth]{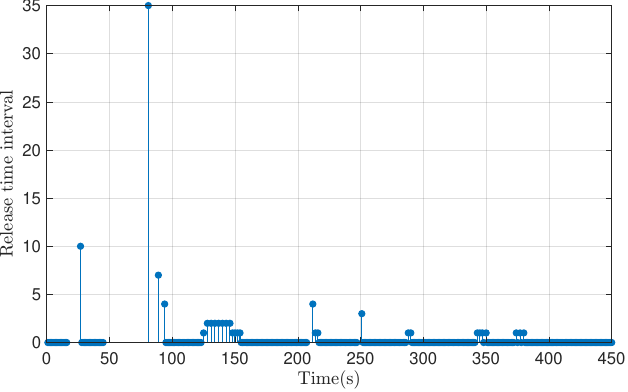}}
    \caption{  {The performance of observer-based ETC under $s_{\rm a}=15\text{m}$}}
    \label{event-control_obs}
\end{figure}

We first test the open-loop system performance. The open-loop results of HVs and AVs are shown in Fig.~\ref{open-loop}. Both traffic density and speed of HVs and AVs oscillate in the whole spatial-temporal domain, leading to the abrupt acceleration of vehicles such that the driving comfort is low. The traffic density and speed do not converge to the equilibrium point. Then, we apply our proposed ETC to the mixed-autonomy traffic system, the closed-loop results are shown in Fig.~\ref{cl_event}. It is observed that the traffic density and speed converge to their equilibrium states exponentially. The traffic density and speed in the spatial-temporal domain reach the equilibrium points. Fig.~\ref{event-control} compares the backstepping controller and the proposed ETC and gives the triggered times in the whole simulation period. It reveals that the ETC is a piecewise-continuous control input, and it almost tracks the backstepping controller when the triggering condition is satisfied. The triggered times is 160 and the total release time is 291s, meaning that the control input remains the same with the last time step until the triggering condition is executed.   {In addition, the traffic density and speed evolution under the observer-based ETC are shown in Fig.\ref{cl_event_obs}, while the comparison of control input and total release time is presented in Fig.\ref{event-control_obs}. The results indicate that, compared to the full-state feedback ETC, the observer-based ETC exhibits slower convergence of traffic density and speed to the equilibrium point. This is due to the additional transient required for the estimated states to approach the true system states. Furthermore, more frequent controller updates are necessary to ensure system stability. The number of triggering times is 358, with a total release time of 93s.}

  {We apply the control input at the downstream of the road section through ramp-metering. The ramp-metering is realized by placing the traffic signal to control the rate whether vehicles enter the main road. By applying ETC to the signal control, the signal does not change at every time step. This reduces the frequency of signal updates, allowing drivers to spend less time monitoring the traffic light and more time focusing on driving. As a result, driver distraction is reduced under longer release time, which contributes to improved safety in the mixed-autonomy traffic system. }


\subsection{Experiments with different spacing polices}
\begin{figure}[!tbp]
    \centering
    \subfloat[  {The closed-loop density and speed evolution of HVs} \label{HVsevent_s16}]{ \includegraphics[width =0.95\linewidth]{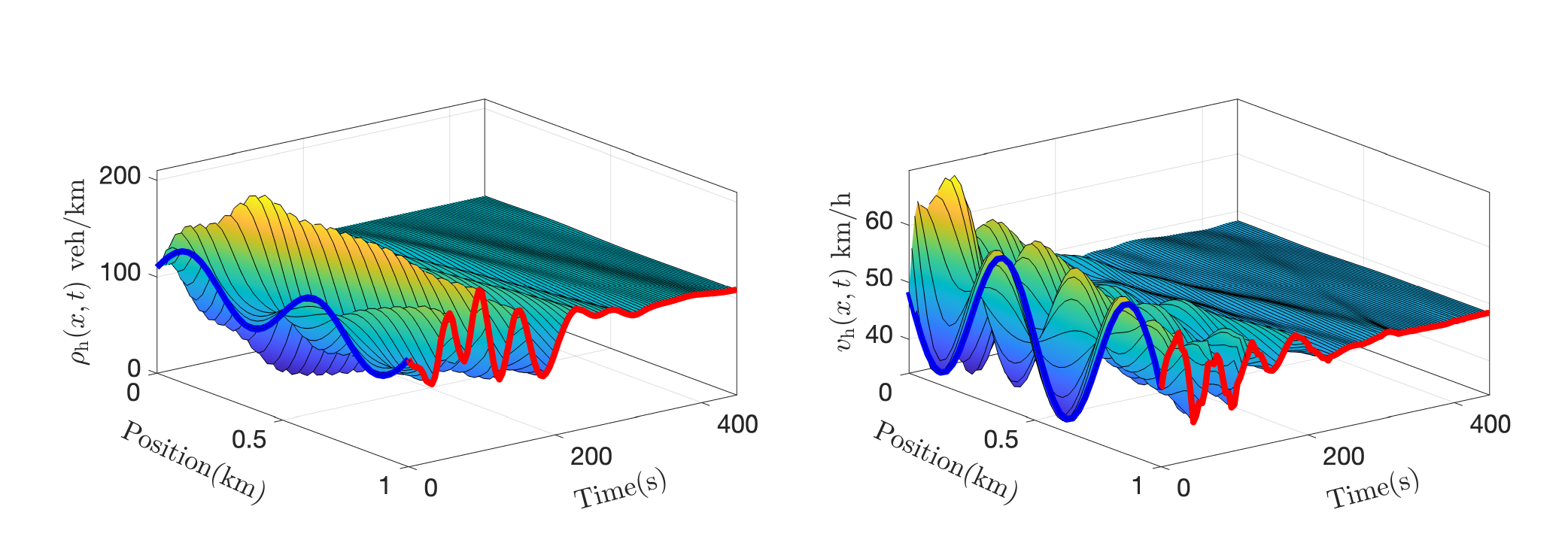}} \\
    \subfloat[  {The closed-loop density and speed evolution of AVs} \label{AVsevent_s16}]{\includegraphics[width =0.95\linewidth]{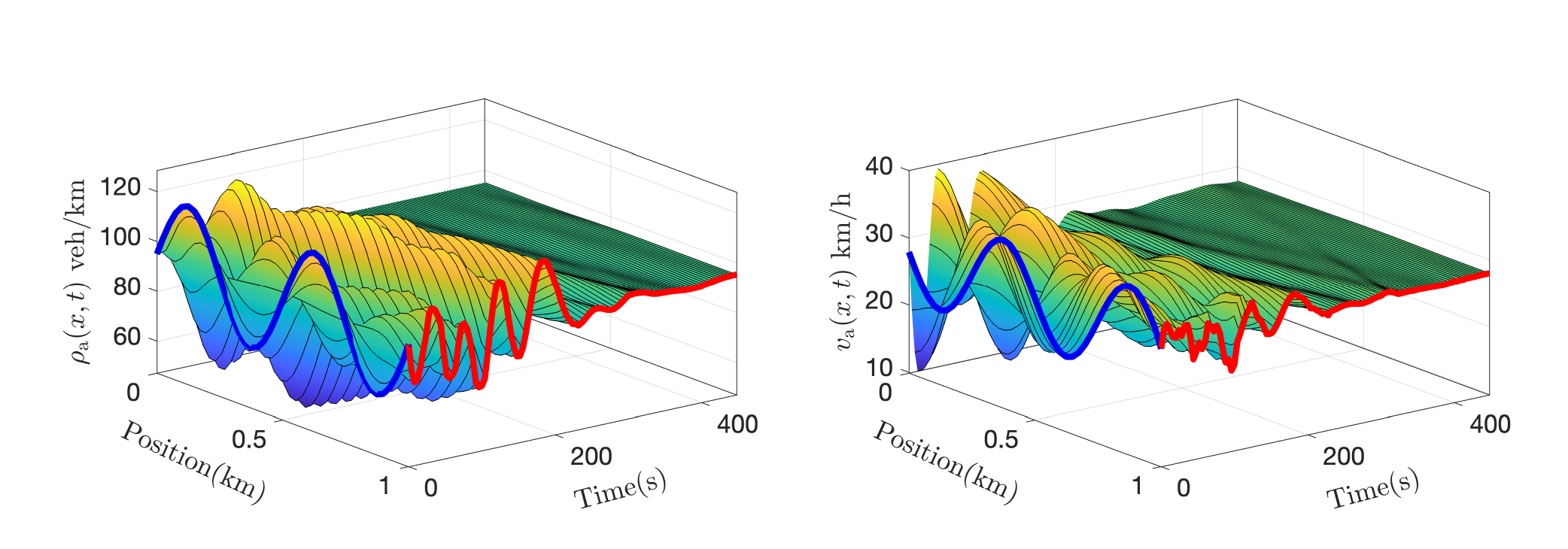}}
    \caption{  {The closed-loop traffic density and speed of mixed-autonomy system under proposed ETC with AV's spacing policy $s_{\rm a}=12\text{m}$}}
    \label{cl_event_s16}
\end{figure}

\begin{figure}[!tbp]
    \centering
    \subfloat[Control input \label{control_compare_s16}]{\includegraphics[width= 0.45\linewidth]{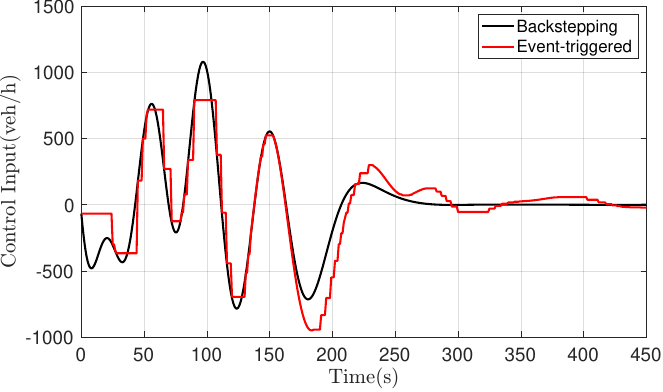}}
    \subfloat[Release instants \label{releaseinterval_s16}]{\includegraphics[width= 0.45\linewidth]{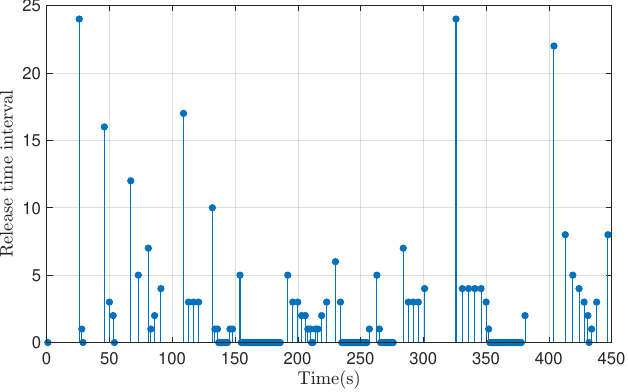}}
    \caption{The performance of ETC under $s_{\rm a}=12\text{m}$}
    \label{event-control_s16}
\end{figure}

In this section, we test the effect of different spacing policies. Taking the $s_{\rm a} = 12 \text{m}$, the traffic density and speed are shown in Fig.~\ref{cl_event_s16}. The other settings keep the same with $s_{\rm a} = 15 \text{m}$, then the different spacing leads to the equilibrium speed as $v^\star_{\rm h} = 48 \text{km/h}$ and $v^\star_{\rm a} =28 \text{km/h}$. It shows that the traffic density and speed also converge to their equilibrium points. The comparison of the backstepping control law and proposed ETC is shown in Fig.~\ref{event-control_s16}(a), while the release instants are shown in Fig.~\ref{event-control_s16}(b). Our proposed ETC stabilizes the system under different spacing policies.

  {In the previous setting, the spacing policy of AVs is still larger than that of HVs. In this section, we explore a more aggressive scenario in which the spacing policy of AVs is smaller than that of HVs. Specifically, the spacing policy for AVs and HVs are set to 8m and 15m, respectively. By adopting this configuration, the corresponding free-flow speeds and maximum area occupancies of AVs and HVs are adjusted accordingly to reflect their altered driving behavior. The density and speed evolution of AVs and HVs are shown in Fig.~\ref{cl_event_agg}. It is observed that the traffic density and speed of HVs and AVs still converge to their equilibrium points at finite time. The comparison of the backstepping controller and proposed ETC is shown in Fig.~\ref{event-control_agg}(a), while the release instants are shown in Fig.~\ref{event-control_agg}(b). Under this aggressive setting, the system achieves a total release time of 349s with only 101 triggering events, indicating a significant reduction in control updates and improved release time.
\begin{figure}[!tbp]
    \centering
    \subfloat[  {The closed-loop density and speed evolution of HVs} \label{HVsevent_agg}]{ \includegraphics[width =0.95\linewidth]{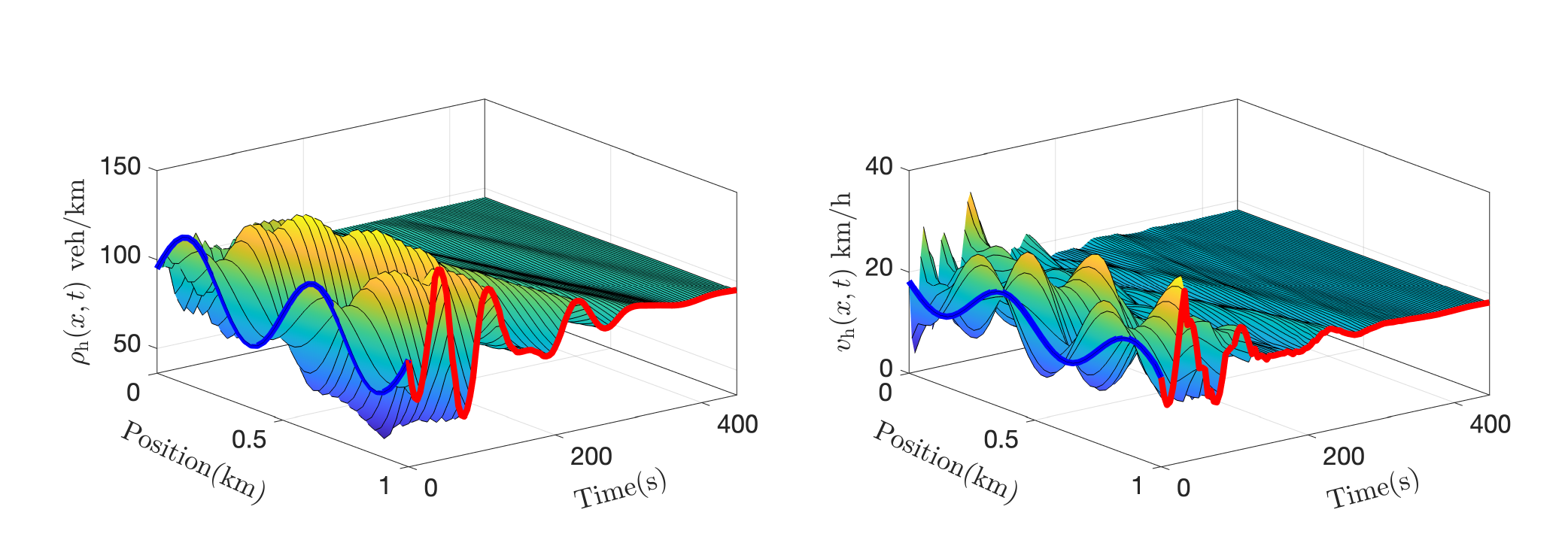}} \\
    \subfloat[  {The closed-loop density and speed evolution of AVs} \label{AVsevent_agg}]{\includegraphics[width =0.95\linewidth]{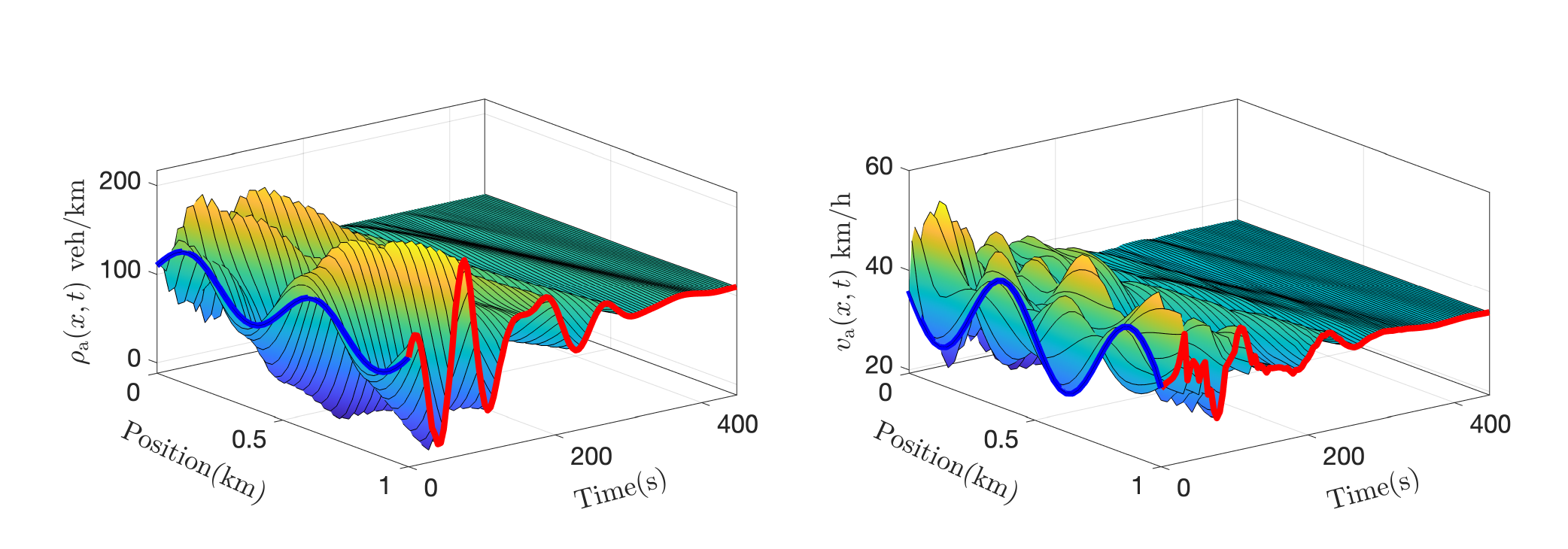}}
    \caption{  {The closed-loop traffic density and speed of mixed-autonomy system under proposed ETC with aggressive AV's spacing policy $s_{\rm a}=8\text{m}$}}
    \label{cl_event_agg}
\end{figure}
\begin{figure}[!tbp]
    \centering
    \subfloat[Control input \label{control_compare_agg}]{\includegraphics[width= 0.45\linewidth]{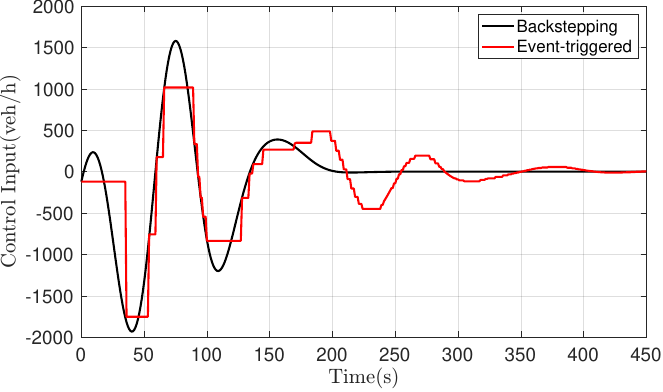}}
    \subfloat[Release instants \label{releaseinterval_agg}]{\includegraphics[width= 0.45\linewidth]{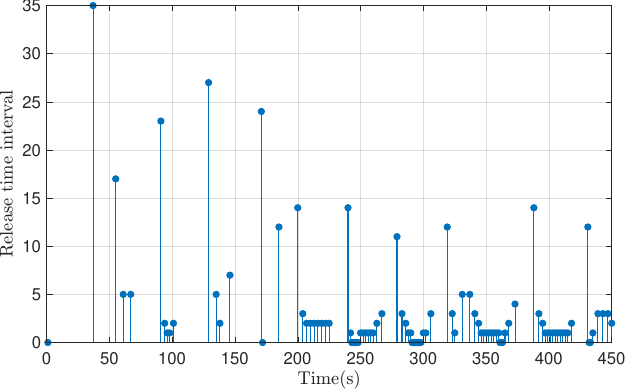}}
    \caption{The performance of ETC under aggressive AV's spacing policy $s_{\rm a}=8\text{m}$}
    \label{event-control_agg}
\end{figure}
}

\subsection{Experiments with different initial traffic conditions}
\begin{figure}[tbp]
    \centering
    \subfloat[  {The traffic density and speed evolution of HVs}]{ \includegraphics[width =0.95\linewidth]{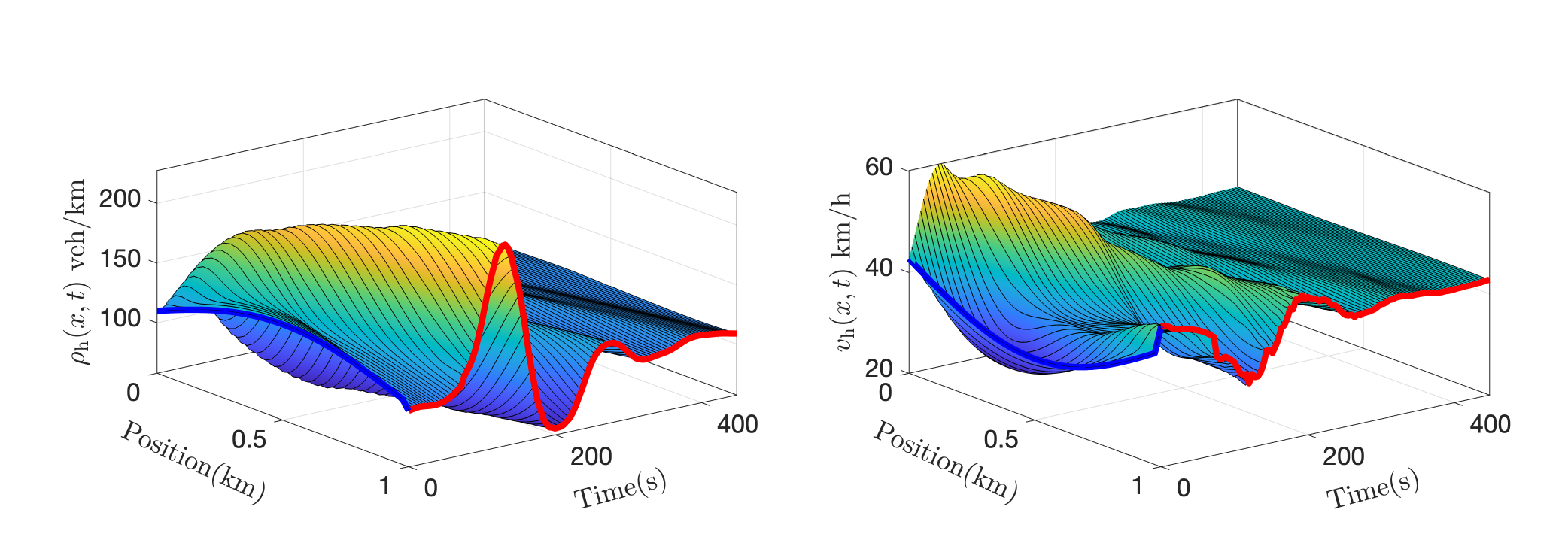}} \\
    \subfloat[  {The traffic density and speed evolution of AVs}]{\includegraphics[width =0.95\linewidth]{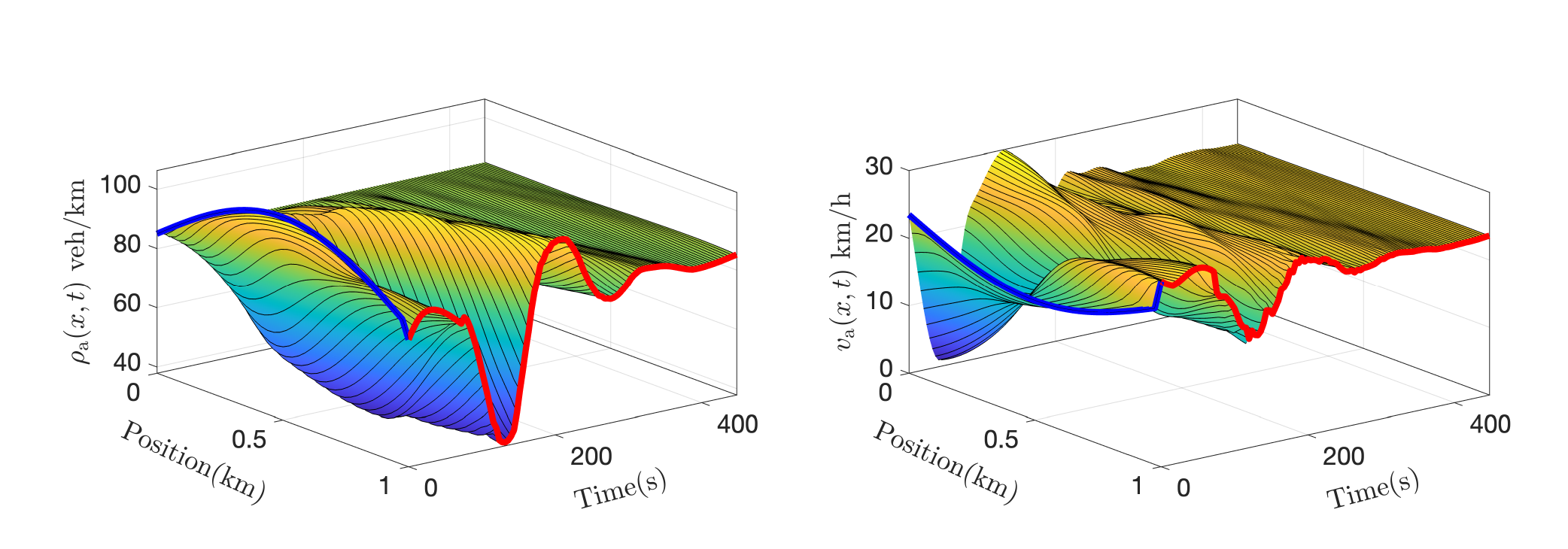}}
    \caption{  {The closed-loop traffic density and speed of mixed-autonomy system under proposed ETC with non-recurrent initial conditions}}
    \label{cl_event_nonini}
\end{figure}

\begin{figure}[tbp]
    \centering
    \subfloat[  {The traffic density and speed evolution of HVs} ]{ \includegraphics[width =0.95\linewidth]{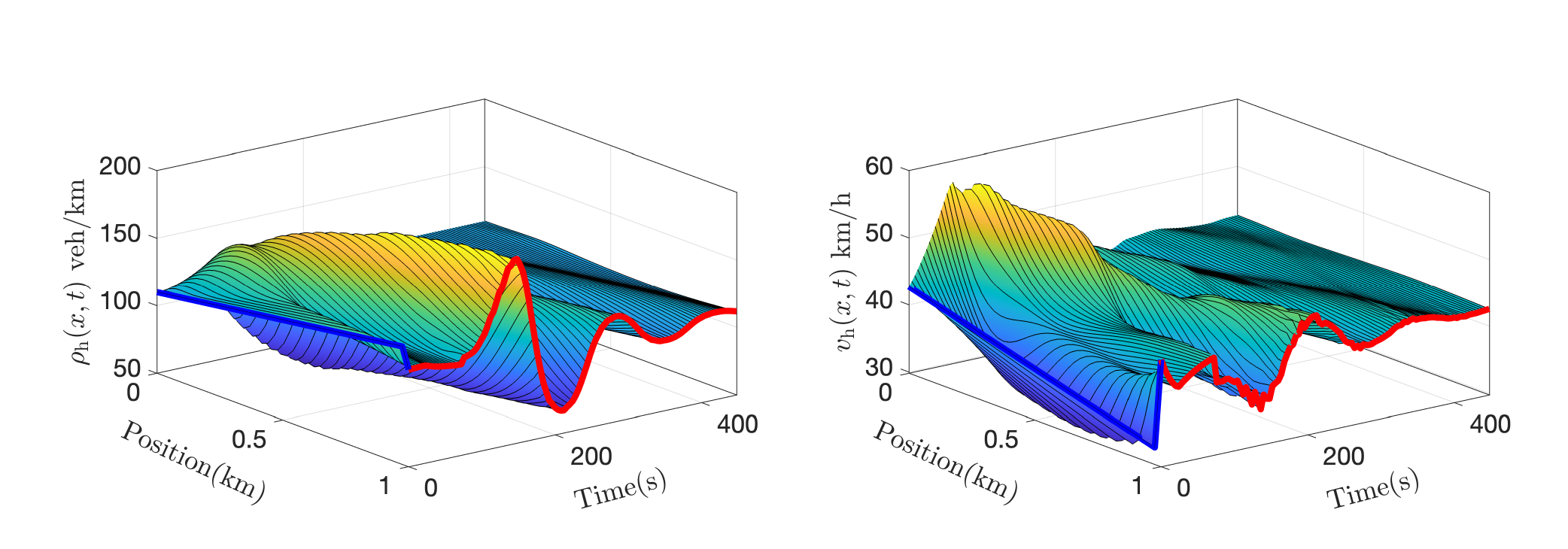}} \\
    \subfloat[  {The traffic density and speed evolution of AVs}]{\includegraphics[width =0.95\linewidth]{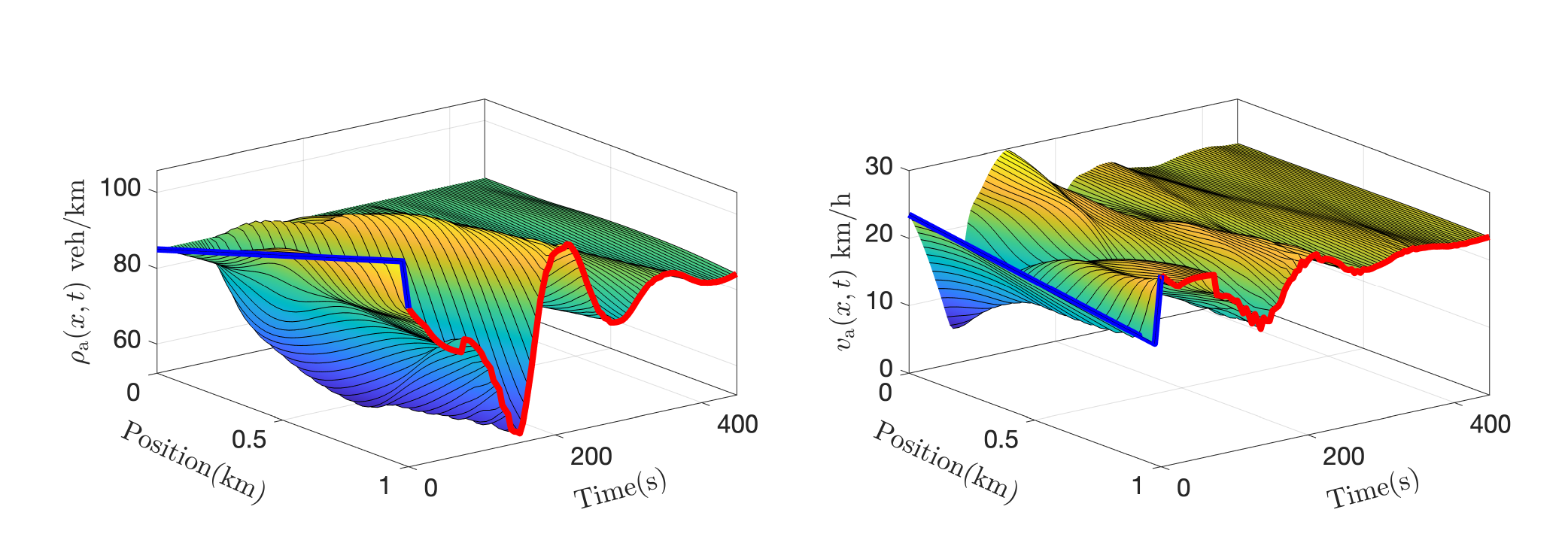}}
    \caption{  {The closed-loop traffic density and speed of mixed-autonomy system under proposed ETC with non-recurrent initial conditions}}
    \label{cl_event_linini}
\end{figure}

\begin{figure}[tbp]
    \centering
    \subfloat[Non-recurrent]{ \includegraphics[width =0.45\linewidth]{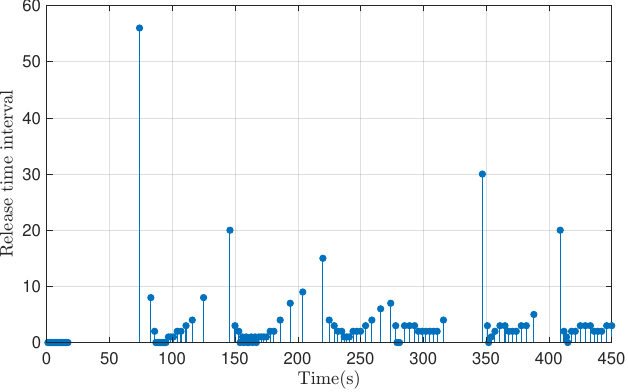}}
    \subfloat[Linear ]{\includegraphics[width =0.45\linewidth]{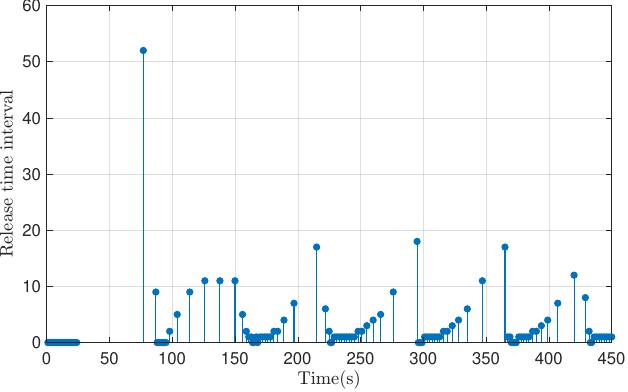}}
    \caption{The release interval with different initial conditions}
    \label{release_interval_ini}
\end{figure}

To further test the robustness of the proposed ETC for the mixed-autonomy traffic system. We take two different initial conditions to test the performance of the ETC. The first is the non-recurrent initial condition, which we select as 
\begin{align}
    \rho_{\rm h}(x,0) &= \rho_{\rm h}^\star + \frac{\rho_{\rm h}^\star }{4} \sin{\left(\frac{\pi x}{L}\right)},
    v_{\rm h}(x,0) = v_{\rm h}^\star - \frac{v_{\rm h}^\star }{4} \sin{\left(\frac{\pi x}{L}\right)},\\
    \rho_{\rm a}(x,0) &= \rho_{\rm a}^\star + \frac{\rho_{\rm a}^\star }{4} \sin{\left(\frac{\pi x}{L}\right)},
    v_{\rm a}(x,0) = v_{\rm a}^\star - \frac{v_{\rm a}^\star }{4} \sin{\left(\frac{\pi x}{L}\right)}.
\end{align}
This condition is related to a sudden deceleration occurring in the middle of the road section, leading to a density wave propagating from upstream to downstream and a speed wave transports from downstream to upstream at the meantime. Therefore, the density will initially increase and then decrease, whereas the speed will decrease and subsequently increase. The results of the non-recurrent initial condition are shown in Fig.~\ref{cl_event_nonini}. The traffic system is still stabilized by the ETC and reaches equilibrium points with small perturbations. Then, we selected the linear initial condition:
\begin{align}
    \rho_{\rm h}(x,0) &= \frac{\rho_{\rm h}^\star }{4} x,  v_{\rm h}(x,0) = -\frac{v_{\rm h}^\star }{4} x,\\
    \rho_{\rm a}(x,0) &= \frac{\rho_{\rm a}^\star }{4} x,  v_{\rm a}(x,0) = -\frac{v_{\rm a}^\star }{4} x.
\end{align}
This condition represents the scenario in which vehicles decelerate downstream due to lane closure. This scenario leads to a decrease in traffic speed and an increase in traffic density due to vehicle accumulation.
The results are shown in Fig.~\ref{cl_event_linini}. It finds that the traffic could also be stabilized under this type of setting. For the non-recurrent case, the system could execute less times to gain a large total release time to improve the driving safety, as shown in Fig.~\ref{release_interval_ini}. The results of both types of initial conditions demonstrate that the proposed ETC effectively stabilizes the traffic system and the traffic states converge to their equilibrium, showing that the proposed ETC is robust to various traffic conditions.

\subsection{ETC under different traffic demands}
\begin{figure}[!tbp]
    \centering
    \subfloat[High]{ \includegraphics[width =0.32\linewidth]{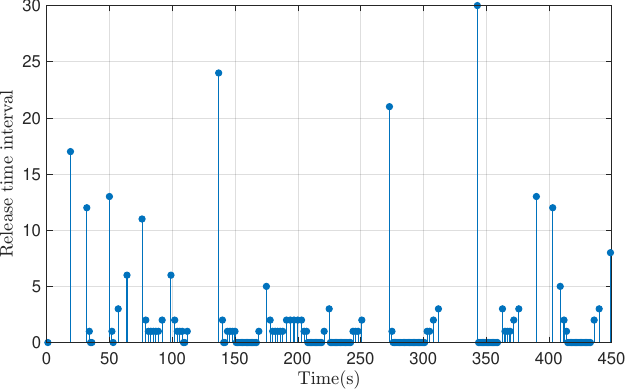}}
    \subfloat[Medium]{\includegraphics[width =0.32\linewidth]{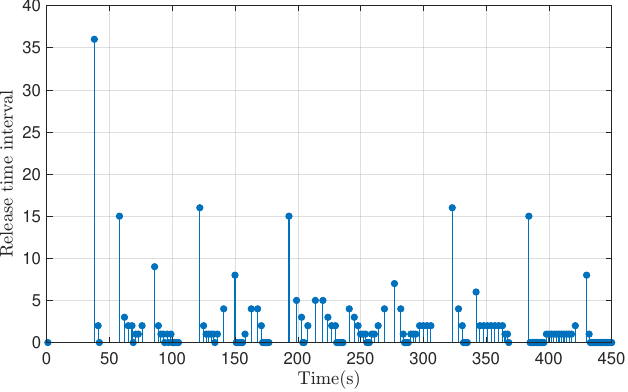}}
    \subfloat[Low]{\includegraphics[width =0.32\linewidth]{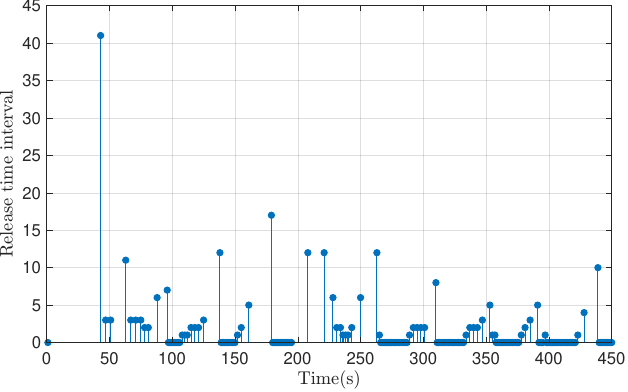}}
    \caption{The release interval with different traffic demands}
    \label{release_interval_demand}
\end{figure}

The proposed ETC has been applied to the various initial conditions to show its effectiveness. In addition to the various initial conditions, the arrival rate of HVs and AVs are varying which leads to different traffic demands on the road. In this section, different traffic demands are applied to the traffic system under the proposed ETC. The different demands are selected as Tab.~\ref{diffdemands}. Three traffic demand levels, high demand, medium demand and low demand, are assigned to test the performance of ETC. It could be concluded that the proposed ETC stabilizes the traffic system in response to different upstream demand, indicating that the ETC is robust to different traffic conditions. The release instant for different demand levels is shown in Fig.~\ref{release_interval_demand}, the controller under the medium demand achieves the longest total release time~(291s) with minimum triggered times~(160) among the three scenarios.
\begin{table}[tbp]
    \centering
    \caption{  {Different traffic demand scenarios containing HV's demand and AV's demand: high demand, medium demand and low demand.}}
    \begin{tabular}{c c c}
    \hline
    \hline
    Total Demand (Veh/h) &  HV Demand (Veh/h)  & AV Demand (Veh/h)\\
    \hline
    \hline
    6510~(High) & 4555 & 1955\\
    5664~(Medium) & 4104 & 1560\\
    4966~(Low) & 3461 & 1505\\
    \hline
    \hline
    \end{tabular}
    \label{diffdemands}
\end{table}

\subsection{Traffic performance evaluation}
We also added three traffic performance indices to compare the performance of the ETC, including fuel consumption, total travel time (TTT) and discomfort value introduced in~\cite{treiber2013traffic}. The definition of the performance indices are:
  {
    \begin{align}
        J_{\text{fuel}} =& \int_0^{T}\int_0^L \max\{0,b_0+ b_1 v(x,t) +b_2v(x,t)a(x,t) \nonumber\\
        &+ b_3v^3(x,t)\} \rho(x,t) dx dt\\
        J_{\text{discom}} =& \int_0^T \int_0^L (a^2(x,t) + a_t^2(x,t))\rho(x,t) dx dt\\
        J_{\text{TTT}} =& \int_0^T \int_0^L\rho(x,t) dx dt
    \end{align}}
  {where coefficients of fuel consumption model are selected as $b_0=25\times10^{-3} \text{1/s}$, $b_1 = 24.5\times 10^{-6}$1/m, $b_2 = 125\times 10^{-6} s^2/m^2$, $b_3 = 32.5\times 10^{-9} s^2/m^3$~\cite[p.485]{treiber2013traffic}.} $a(x,t)$ is the local acceleration $a(x,t) = v_t(x,t) + v(x,t) v_x(x,t)$.
As shown in Tab.~\ref{diffperformance}, the performance indices of each controller and different scenarios are compared with their improvement percentage over the baseline open-loop system. It reveals that the backstepping controller and ETC could significantly improve the driving comfort compared with the open-loop system, leading to a reduction of the discomfort value. Compared with the backstepping controller, the ETC brings discomfort to drivers due to the piece-wise constant control input induced by the event-triggered mechanism,  while the driving safety is improved and the computational resource is saved.   {It is also observed that the controller significantly improves fuel consumption and total travel time under non-recurrent and linear initial conditions, as these scenarios are less complex compared to the stop-and-go traffic congestion scenario. The proposed ETC also achieves a longer release time in these two cases.}

\begin{table}[tbp]
    \centering
    \caption{  {Traffic performance of the mixed-autonomy traffic system in different spacing policies of AVs, different initial conditions and different traffic demands}}
    \begin{tabular}{c c c c c }
    \hline
    \hline
    Scenario &  $J_{\text{fuel}}$   & $J_{\text{discom}}$  & $J_{\text{TTT}}$ &  \makecell{ Release \\time(s)} \\
    \hline
    \hline
    Backstepping(15) & -0.57\% & -48.86\% & -0.52\% & 0 \\
    ETC(15) & -0.39\% & -41.07\% & -0.34\% & {291} \\
    \hline
    \hline
    \multicolumn{5}{c}{\textbf{Different spacing policy of AVs}}\\
    \hline
    \hline
    Backstepping(12) & -0.14\% & -74.27\% & -0.03\% & 0 \\
    ETC(12) & -0.05\% & -71.17\% & -0.17\% & {284} \\
    \hline
    \hline
    \multicolumn{5}{c}{\textbf{Aggressive spacing policy of AVs}}\\
    \hline
    \hline
    Backstepping(8) & -0.83\% & -73.63\% & -0.88\% & 0 \\
    ETC(8) & -0.86\% & -26.46\% & -0.90\% & {349} \\
    \hline
    \hline
    \multicolumn{5}{c}{\textbf{Different initial conditions}}\\
    \hline
    \hline
    Non-recurrent & -10.82\% & -89.01\% & -11.37\% & {335} \\
    Linear & -9.27\% & -74.68\% & -9.64\% & 324 \\
    \hline
    \hline
    \multicolumn{5}{c}{\textbf{Different traffic demands}}\\
    \hline
    \hline
    High  & -0.30\% & -57.96\% & -0.22\% & {265} \\
    
    Medium  & -0.39\% & -41.07\% & -0.34\% & {291} \\
    
    Low  & -0.75\% & -15.54\% & -0.76\% & 252 \\
    \hline
    \hline
    \end{tabular}
    \label{diffperformance}
\end{table}
Another traffic performance index is the total delay(TD) caused by the traffic congestion. Smaller delay indicates the high efficiency of the mixed-autonomy traffic system, the definition of the delay is defined as~\cite{papageorgiou2002freeway}:
\begin{align}
    \text{TD}_i = \text{TTT}_i - \frac{\text{TMT}_i}{V_{i}}, \quad i \in \{ \text{HVs}, \text{AVs} \},
\end{align}
where TTT is the total travel time, and TMT is the total miles traveled. $V_i$ is the maximum traffic speed of HVs and AVs. We calculate the total delay for the HVs and AVs in different scenarios, as shown in Tab.~\ref{totaldelay}. Compared with the open-loop system, the closed-loop system performs lower total delays under both the backstepping method and the proposed ETC. The proposed ETC can reduce delays more effectively than the backstepping controller in aggressive-spacing AV settings, demonstrating its superior ability to mitigate traffic congestion. It is also observed that the delay in the linear initial condition outperforms the non-recurrent initial condition, while under different traffic demands, the proposed ETC has smallest delay under high traffic demands.
\begin{table}[tbp]
    \centering
    \caption{  {Total delay of the mixed-autonomy traffic system under different spacing policies of AVs, initial conditions, and traffic demands}}
    \begin{tabular}{c c c c}
    \hline
    \hline
    Scenario &  HVs delay   & AVs delay  & Total delay \\
    \hline
    \hline
    Backstepping(15) & -0.12\% & -0.68\% & -0.40\% \\
    ETC(15)  & -0.46\% & -0.56\% & -0.52\%\\
    \hline
    \hline
    \multicolumn{4}{c}{\textbf{Different spacing policy of AVs}}\\
    \hline
    \hline
    Backstepping(12) & -3.80\% & +2.28\% & -0.65\% \\
    ETC(12) & -4.25\% & +1.75\% & -1.14\%\\
    \hline
    \hline
    \multicolumn{4}{c}{\textbf{Aggressive spacing policy of AVs}}\\
    \hline
    \hline
    Backstepping(8) & +2.20\% & -19.58\% & -7.28\% \\
    ETC(8) & +2.24\% & -18.67\% & -6.85\% \\
    \hline
    \hline
    \multicolumn{4}{c}{\textbf{Different initial conditions}}\\
    \hline
    \hline
    Non-recurrent & -33.41\% & -10.21\% & -23.98\% \\
    Linear & -27.16\% & -10.34\% & -19.80\%\\
    \hline
    \hline
    \multicolumn{4}{c}{\textbf{Different traffic demands}}\\
    \hline
    \hline
    High  & -1.54\% & -0.17\% & -0.86\% \\
    Medium  & -0.46\% & -0.56\% & -0.52\% \\
    Low  & -1.41\% & -1.28\% & -1.34\% \\
    \hline
    \hline
    \end{tabular}
    \label{totaldelay}
\end{table}

\subsection{Experiments with different AV penetration rates}
We further test the effectiveness of the event-triggered mechanism under different penetration rates of AVs. The penetration rate of AVs is defined as the total number of AVs on the studied road section. We keep the equilibrium inflow to be the same under the different equilibrium density of AVs and HVs and test the different penetration rates of AVs from $44\%$ to $53\%$, the results are shown in Tab.~\ref{diffpene}. The total release time is 206 s with 245 triggering times at $44\%$, while the total release time grows to 299 s with 152 triggering times at $53\%$. It is revealed that the system would execute more times under the low penetration rate to maintain the stability of the system and drive the system states to equilibrium states, while it tends to execute less time to reduce the computational burden and gain more release time under higher penetration rates of AVs. Due to the conservative driving policy of AVs, the system becomes more stable  under high penetration rates that the event-triggering mechanism executes less.
\begin{table}[!tbp]
    \centering
    \caption{The effect of AV penetration rates on the triggering mechanism}
    \begin{tabular}{c c c}
    \hline
    \hline
    Penetration rate &  \makecell{Total \\release time(s)}   &  Triggering times \\
    \hline
    \hline
    44\% & 206 & 245\\ 
    47\% & 287 & 161\\ 
    50\% & 292 & 158\\ 
    53\% & 299 & 152\\ 
    \hline
    \hline
    \end{tabular}
    \label{diffpene}
\end{table}

\subsection{Sensitivity analysis of design parameters}
\begin{figure}[tbp]
    \centering
    \subfloat[$\mu$]{ \includegraphics[width =0.32\linewidth]{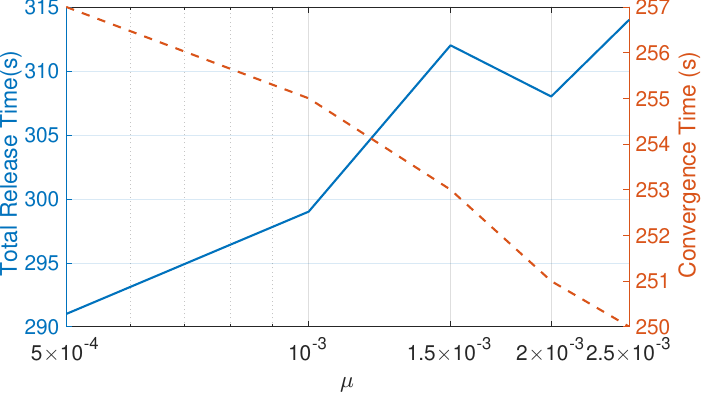}}
    \subfloat[$\sigma$ ]{\includegraphics[width =0.32\linewidth]{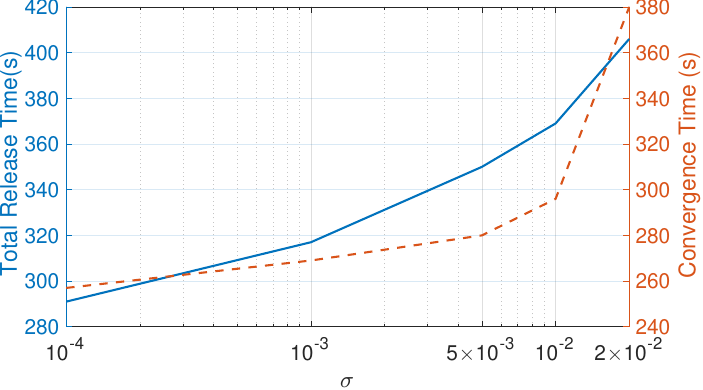}}
    \subfloat[$\zeta$ ]{\includegraphics[width =0.32\linewidth]{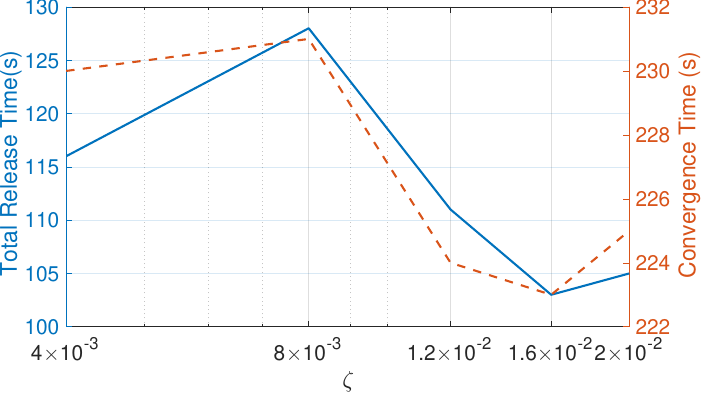}}
    \caption{Sensitivity analysis for design parameters $\mu$, $\sigma$, and $\zeta$. The total release time and convergence time of Lyapunov functional are selected as performance indices.}
    \label{sensitive_analysis}
\end{figure}

  {In the previous section, we have proved that the mixed-autonomy traffic system achieves exponentially convergence. The designed parameters we use are selected by~\eqref{select1}-\eqref{select2} and~\eqref{select3}-\eqref{select4}. In this section, we conduct sensitive analysis for the designed parameters $\mu$, $\sigma$, and $\zeta$. Two performance measure, total release time and convergence time of Lyapunov functional, are selected to test the system performance under different design parameters. The convergence time of Lyapunov functional denotes the time when the value of the Lyapunov functional smaller than certain threshold. Results for different values of  $\mu$, $\sigma$, and $\zeta$ are shown in Fig.~\ref{sensitive_analysis}. It is observed that a larger $\mu$ leads to a longer release time and a shorter convergence time of Lyapunov functional, while a larger~$\sigma$ leads to both longer release time and convergence time. This is consistent with our theoretical results in~\eqref{estimate4ori} that a larger $\sigma$ would make the system converge slower. For the sensitive analysis of $\zeta$, it is related to triggering condition in Definition~\ref{defdynamic}. The result shows that the release time and the convergence time exhibit the same trend under the different value of $\zeta$.} 

\section{Conclusion and Future Work}\label{sec6}
In this paper, we proposed an ETC strategy to mitigate stop-and-go traffic congestion in mixed-autonomy systems consisting of both AVs and HVs. The propagation of densities and speeds of HVs and AVs in congested traffic was described by a hyperbolic 4 $\times$ 4 PDEs and boundary control input was considered. The ETC was constructed based on continuous-time control laws derived through backstepping control design to reduce the computational burden. Theoretical analysis was conducted to prove the effectiveness of the proposed ETC and the Zeno phenomenon was proved to be avoided using the Lyapunov method. Finally, extensive simulations were applied to verify theoretical results and validate the proposed ETC under various traffic scenarios, including different traffic demands, AV penetration rates, and different spacing policies. It was shown that higher AV penetration rates lead to longer release time, giving the possibility of AVs in mitigating traffic congestion under the proposed ETC.

  {Although we have formulated the observer-based ETC, a rigorous analysis for the mixed-autonomy traffic system to prove the Zeno-free behavior and the finite-time convergence is still needed. 
The total release time is reduced in observer-based ETC due to the requirement of ensuring both system stability and convergence of traffic states. This necessitates a higher frequency of triggering events, thereby leading to less release time. The traffic performance (TTT, driving discomfort value, fuel consumption) may decrease compared with the full-state feedback controller due to the states estimation error caused by the observer. In addition, it is also possible to utilize machine learning-based methods such as physics-informed neural networks(PINNs)~\cite{karniadakis_physics-informed_2021,zhao2023observer} and neural operators(NO)~\cite{zhang2024operator,lu2021learning,kovachki2023neural} for traffic state estimation. These approaches enable the reconstruction of full traffic state information, which can subsequently be used to compute the backstepping control law. Future work will focus on the theoretical analysis of the observer-based ETC.} 

  {Another limitation of this work lies in computational load of monitoring the triggering condition at each time-step in the whole simulation period, which also imposes significant pressure on the traffic management system. A possible solution is to adopt periodic ETC and self-triggered ETC designs~\cite{zhang2024performance,SOMATHILAKE2025}, which avoid the need for continuous sensing of the triggering condition and Lyapunov functionals.}






\bibliographystyle{ieeetr}
\bibliography{reference}
\flushend
\begin{IEEEbiography}[{\includegraphics[width=1in,height=1.25in,clip,keepaspectratio]{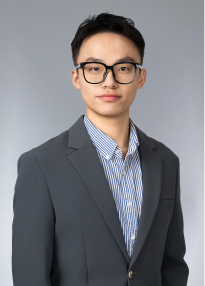}}]{Yihuai Zhang} (Graduate Student Member, IEEE) received his B.E. degree in Vehicle Engineering from Southwest University in 2019, and his M.S. degree in Vehicle Engineering from South China University of Technology in 2022. He is currently a Ph.D. student in the Thrust of Intelligent Transportation at the Hong Kong University of Science and Technology (Guangzhou). His doctoral research focuses on distributed parameter systems, learning and control for dynamical systems, and their applications in transportation systems.
\end{IEEEbiography}
\begin{IEEEbiography}
[{\includegraphics[width=1in,height=1.25in,clip,keepaspectratio]{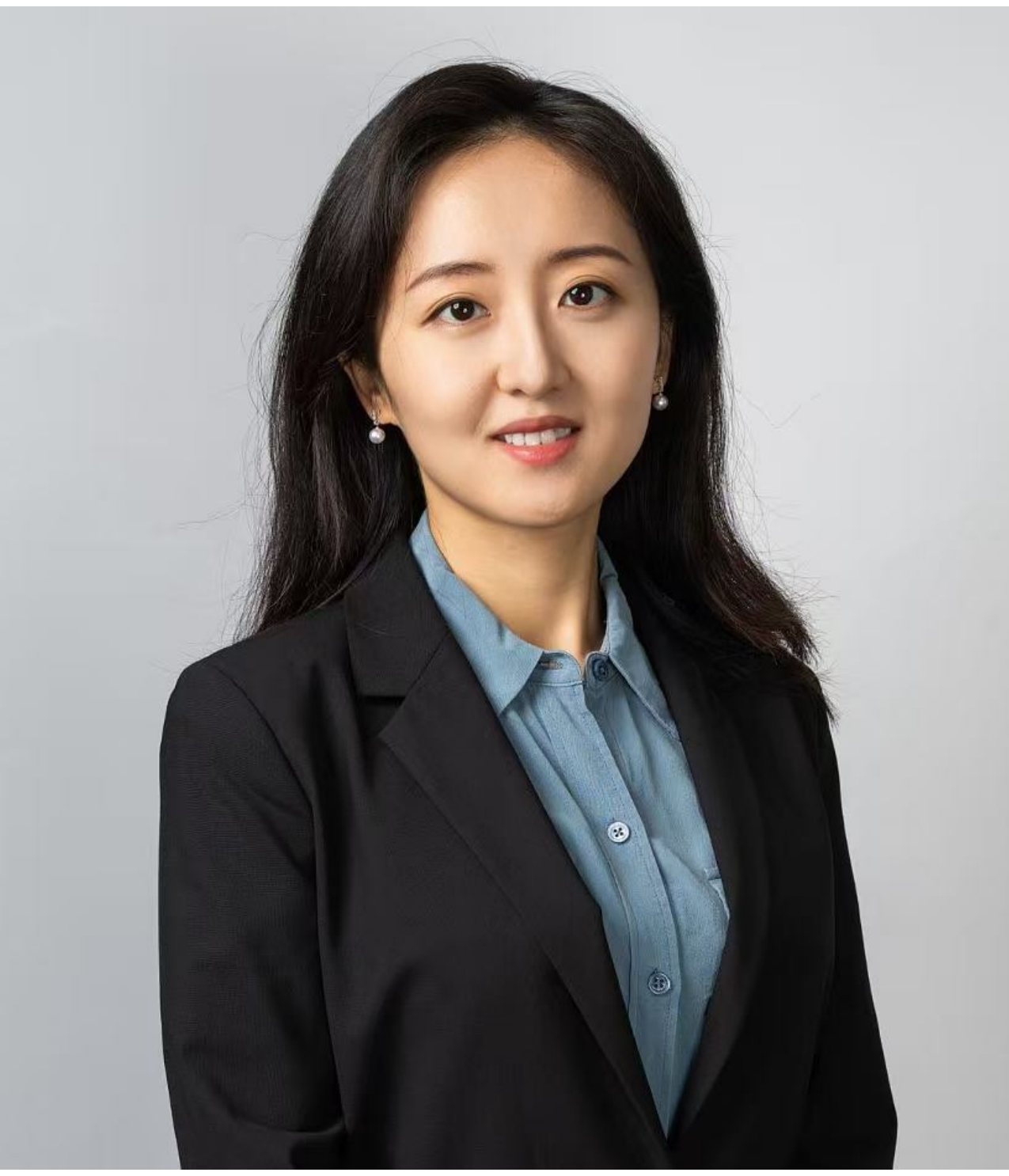}}]{Huan Yu} (Senior Member, IEEE) is an Assistant Professor in the Intelligent Transportation Thrust at the Hong Kong University of Science and Technology (Guangzhou). Yu received her B.Eng. degree in Aerospace Engineering from Northwestern Polytechnical University, and the M.Sc. and Ph.D. degrees in Aerospace Engineering from the Department of Mechanical and Aerospace Engineering, University of California, San Diego. She was a visiting scholar at University of California, Berkeley and Massachusetts Institute of Technology. She is broadly interested in control theory, optimization, and machine learning methodologies and their applications in intelligent transportation systems.
\end{IEEEbiography}

\end{document}